\documentclass[prb,twocolumn,superscriptaddress, preprintnumbers,letterpaper,twoside]{revtex4}

\usepackage{times}
\usepackage{amsmath}
\usepackage{amsthm}
\usepackage{amssymb}
\usepackage{amsbsy}
\usepackage{amsfonts}
\usepackage{bm}
\usepackage{graphicx}
\usepackage{multirow}

\renewcommand{\vec}[1]{{\mathbf #1}}

\renewcommand{\ol}[1]{\overline{#1}}

\newcommand{\comments}[1]{}

\newcommand{\nc}{\newcommand}

\def\Aut{\mathrm{Aut}}

\def\Tr{\mathrm{Tr}}

\def\GL{\mathrm{GL}}

\makeatletter
\let\O\@undefined
\def\O{\mathrm{O}}
\def\SO{\mathrm{SO}}

\def\U{\mathrm{U}}
\def\SU{\mathrm{SU}}

\DeclareMathOperator{\Spec}{Spec}

\def\smfrac#1#2{{ \mbox{$\frac{#1}{#2}$} }}

\def\vev#1{\langle #1 \rangle}

\def\pmat#1{{\begin{pmatrix}#1\end{pmatrix}}}

\def\smat#1{{\bigl(\begin{smallmatrix}#1\end{smallmatrix}\bigr)}}

\def\leg#1#2{{#1 \overwithdelims () #2}}

\nc{\CA}{{\cal A}} \nc{\CB}{{\cal B}} \nc{\CC}{{\cal C}}
\nc{\CD}{{\cal D}} \nc{\CE}{{\cal E}} \nc{\CF}{{\cal F}}
\nc{\CG}{{\cal G}} \nc{\CH}{{\cal H}} \nc{\CI}{{\cal I}}
\nc{\CJ}{{\cal J}} \nc{\CK}{{\cal K}} \nc{\CL}{{\cal L}}
\nc{\CM}{{\cal M}} \nc{\CN}{{\cal N}} \nc{\CO}{{\cal O}}
\nc{\CP}{{\cal P}} \nc{\CQ}{{\cal Q}} \nc{\CR}{{\cal R}}
\nc{\CS}{{\cal S}} \nc{\CT}{{\cal T}} \nc{\CU}{{\cal U}}
\nc{\CV}{{\cal V}} \nc{\CW}{{\cal W}} \nc{\CX}{{\cal X}}
\nc{\CY}{{\cal Y}} \nc{\CZ}{{\cal Z}}

\nc{\bA}{\mathbb{A}} \nc{\bB}{\mathbb{B}} \nc{\bC}{\mathbb{C}}
\nc{\bD}{\mathbb{D}} \nc{\bE}{\mathbb{E}} \nc{\bF}{\mathbb{F}}
\nc{\bG}{\mathbb{G}} \nc{\bH}{\mathbb{H}} \nc{\bI}{\mathbb{I}}
\nc{\bJ}{\mathbb{J}} \nc{\bK}{\mathbb{K}} \nc{\bL}{\mathbb{L}}
\nc{\bM}{\mathbb{M}} \nc{\bN}{\mathbb{N}} \nc{\bO}{\mathbb{O}}
\nc{\bP}{\mathbb{P}} \nc{\bQ}{\mathbb{Q}} \nc{\bR}{\mathbb{R}}
\nc{\bS}{\mathbb{S}} \nc{\bT}{\mathbb{T}} \nc{\bU}{\mathbb{U}}
\nc{\bV}{\mathbb{V}} \nc{\bW}{\mathbb{W}} \nc{\bX}{\mathbb{X}}
\nc{\bZ}{\mathbb{Z}}

\nc{\BA}{\mathbf{A}} \nc{\BB}{\mathbf{B}} \nc{\BC}{\mathbf{C}}
\nc{\BD}{\mathbf{D}} \nc{\BE}{\mathbf{E}} \nc{\BF}{\mathbf{F}}

\nc{\BG}{\mathbf{G}} \nc{\BH}{\mathbf{H}} \nc{\BI}{\mathbf{I}}
\nc{\BJ}{\mathbf{J}} \nc{\BK}{\mathbf{K}} \nc{\BL}{\mathbf{L}}
\nc{\BM}{\mathbf{M}} \nc{\BN}{\mathbf{N}} \nc{\BO}{\mathbf{O}}
\nc{\BP}{\mathbf{P}} \nc{\BQ}{\mathbf{Q}} \nc{\BR}{\mathbf{R}}
\nc{\BS}{\mathbf{S}} \nc{\BT}{\mathbf{T}} \nc{\BU}{\mathbf{U}}
\nc{\BV}{\mathbf{V}} \nc{\BW}{\mathbf{W}} \nc{\BX}{\mathbf{X}}
\nc{\BY}{\mathbf{Y}} \nc{\BZ}{\mathbf{Z}}

\nc{\msA}{\mathscr{A}} \nc{\msB}{\mathscr{B}} \nc{\msC}{\mathscr{C}}
\nc{\msD}{\mathscr{D}} \nc{\msE}{\mathscr{E}} \nc{\msF}{\mathscr{F}}
\nc{\msG}{\mathscr{G}} \nc{\msH}{\mathscr{H}} \nc{\msI}{\mathscr{I}}
\nc{\msJ}{\mathscr{J}} \nc{\msK}{\mathscr{K}} \nc{\msL}{\mathscr{L}}
\nc{\msM}{\mathscr{M}} \nc{\msN}{\mathscr{N}} \nc{\msO}{\mathscr{O}}
\nc{\msP}{\mathscr{P}} \nc{\msQ}{\mathscr{Q}} \nc{\msR}{\mathscr{R}}
\nc{\msS}{\mathscr{S}} \nc{\msT}{\mathscr{T}} \nc{\msU}{\mathscr{U}}
\nc{\msV}{\mathscr{V}} \nc{\msX}{\mathscr{X}} \nc{\msW}{\mathscr{W}}
\nc{\msY}{\mathscr{Y}} \nc{\msZ}{\mathscr{Z}}

\nc{\mfa}{{\mathfrak a}} \nc{\mfb}{{\mathfrak b}} \nc{\mfc}{{\mathfrak c}}
\nc{\mfd}{{\mathfrak d}} \nc{\mfe}{{\mathfrak e}} \nc{\mff}{{\mathfrak f}}
\nc{\mfg}{{\mathfrak g}} \nc{\mfh}{{\mathfrak h}} \nc{\mfi}{{\mathfrak i}}
\nc{\mfj}{{\mathfrak j}} \nc{\mfk}{{\mathfrak k}} \nc{\mfl}{{\mathfrak l}}
\nc{\mfm}{{\mathfrak m}} \nc{\mfn}{{\mathfrak n}} \nc{\mfo}{{\mathfrak o}}
\nc{\mfp}{{\mathfrak p}} \nc{\mfq}{{\mathfrak q}} \nc{\mfr}{{\mathfrak r}}
\nc{\mfs}{{\mathfrak s}} \nc{\mft}{{\mathfrak t}} \nc{\mfu}{{\mathfrak u}}
\nc{\mfv}{{\mathfrak v}} \nc{\mfw}{{\mathfrak w}} \nc{\mfx}{{\mathfrak x}}
\nc{\mfy}{{\mathfrak y}} \nc{\mfz}{{\mathfrak z}}

\nc{\mfA}{{\mathfrak A}} \nc{\mfB}{{\mathfrak B}} \nc{\mfC}{{\mathfrak C}}
\nc{\mfD}{{\mathfrak D}} \nc{\mfE}{{\mathfrak E}} \nc{\mfF}{{\mathfrak F}}
\nc{\mfG}{{\mathfrak G}} \nc{\mfH}{{\mathfrak H}} \nc{\mfI}{{\mathfrak I}}
\nc{\mfJ}{{\mathfrak J}} \nc{\mfK}{{\mathfrak K}} \nc{\mfL}{{\mathfrak L}}
\nc{\mfM}{{\mathfrak M}} \nc{\mfN}{{\mathfrak N}} \nc{\mfO}{{\mathfrak O}}
\nc{\mfP}{{\mathfrak P}} \nc{\mfQ}{{\mathfrak Q}} \nc{\mfR}{{\mathfrak R}}
\nc{\mfS}{{\mathfrak S}} \nc{\mfT}{{\mathfrak T}} \nc{\mfU}{{\mathfrak U}}
\nc{\mfV}{{\mathfrak V}} \nc{\mfW}{{\mathfrak W}} \nc{\mfX}{{\mathfrak X}}
\nc{\mfY}{{\mathfrak Y}} \nc{\mfZ}{{\mathfrak Z}}

\nc{\ep}{\epsilon}
\nc{\Om}{\Omega}
\nc{\wt}[1]{\widetilde{#1}}

\def\al{\alpha}

\def\sig{\sigma}

\begin{document}


\title{Bulk-Edge Correspondence in $2+1$-Dimensional Abelian Topological Phases}

\author{Jennifer Cano}
\affiliation{Department of Physics, University of California, Santa Barbara,
California 93106, USA}
\author{Meng Cheng}
\affiliation{Microsoft Research, Station Q, Elings Hall,
University of California, Santa Barbara, California 93106-6105, USA}
\author{Michael Mulligan}
\affiliation{Microsoft Research, Station Q, Elings Hall,
University of California, Santa Barbara, California 93106-6105, USA}
\author{Chetan Nayak}
\affiliation{Microsoft Research, Station Q, Elings Hall,
University of California, Santa Barbara, California 93106-6105, USA}
\affiliation{Department of Physics, University of California, Santa Barbara, California 93106, USA}
\author{Eugeniu Plamadeala}
\affiliation{Department of Physics, University of California, Santa Barbara,
California 93106, USA}
\author{Jon Yard}
\affiliation{Microsoft Research, Station Q, Elings Hall,
University of California, Santa Barbara, California 93106-6105, USA}

\begin{abstract}
The same bulk two-dimensional topological phase can have multiple distinct, fully-chiral
edge phases. We show that this can occur in the integer quantum Hall states at $\nu=8$ and $12$,
with experimentally-testable consequences. We show that this can occur in Abelian fractional
quantum Hall states as well, with the simplest examples being at $\nu=8/7, 12/11, 8/15, 16/5$.
We give a general criterion for the existence of multiple distinct chiral edge phases for the same bulk phase
and discuss experimental consequences.
Edge phases correspond to lattices while bulk phases correspond to genera of lattices.
Since there are typically multiple lattices in a genus, the bulk-edge correspondence is typically one-to-many;
there are usually many stable fully chiral edge phases corresponding to the same bulk.
We explain these correspondences using the theory of
integral quadratic forms. We show that fermionic systems can have edge phases with only bosonic
low-energy excitations and discuss a fermionic generalization of the relation between bulk topological spins and the central charge.
The latter follows from our demonstration that every fermionic topological phase can be represented as a bosonic topological phase,
together with some number of filled Landau levels.
Our analysis shows that every Abelian topological phase can be decomposed into a tensor product of theories associated
with prime numbers $p$ in which every quasiparticle has a topological spin that is a $p^n$-th root of
unity for some $n$. It also leads to a simple demonstration that all Abelian topological phases can be
represented by $\U(1)^N$ Chern-Simons theory parameterized by a K-matrix.
\end{abstract}

\maketitle

\section{Introduction}

In the limit of vanishing electron-electron interactions, the edge excitations of an integer quantum Hall state form a multi-channel chiral Fermi liquid.
These excitations are stable with respect to weak interactions by their chirality \cite{Wengapless}. However, the Coulomb energy in observed integer quantum Hall states is larger than
the energy of the lowest gapped edge excitation. Therefore, interactions are not weak in these experiments, and we must consider whether interactions with gapped
unprotected non-chiral excitations can alter the nature of
the gapless protected chiral edge excitations of an integer quantum Hall state even when the bulk is unaffected.\footnote{In fact, the Coulomb energy is often larger than the bulk cyclotron energy, too,
so it is not a given that the bulk state is in the same universality class as the non-interacting integer quantum Hall state, but we will assume that this is true in this paper.}

In this paper, we show that sufficiently strong interactions can drive the edge of an integer quantum Hall state with $\nu\geq 8$ into a different phase in which the edge excitations form a multi-channel chiral {\it Luttinger liquid} while the bulk remains adiabatically connected to an integer quantum Hall state of non-interacting electrons.
This chiral Luttinger liquid is also stable against all weak perturbations, but it is not adiabatically connected to the edge of an integer quantum Hall state of non-interacting electrons
even though the bulk of the system is.
For $\nu\geq 12$, there are several possible such stable chiral
edge phases corresponding to the same bulk phase.
The edge excitations of many fractional quantum Hall states, such as the principal Jain series with $\nu=\frac{n}{2pn+1}$ form a multi-channel chiral Luttinger liquid, which is stable against weak perturbations due to its chirality. We show that such edges can also be subject to reconstruction
into a different chiral Luttinger liquid as a result of strong interactions with gapped unprotected
excitations at the edge. The new chiral Luttinger liquid is also stable against all weak perturbations.

A similar phenomenon was recently analyzed in the context of bosonic analogues of integer
quantum Hall states\cite{Plamadeala13}.
Without symmetry, integer quantum Hall states of bosons that only
support bosonic excitations in the bulk, not anyons, occur only when the
chiral central charge, $c_- = c_R - c_L$, the difference between the number of right- and left-moving edge modes,
is a multiple of eight (or, equivalently, when the thermal Hall conductance is
$\kappa_{xy} = {c_-}\frac{\pi^2 k_B^2 T} {3 h}$ with $c_-=8k$ for integers $k$).\cite{Kitaev11}
There is a unique \citep{Serre73,Milnor73} bulk state for each possible value of $c_- =8k$, but there are many possible chiral edge phases when the chiral central charge is greater than $8$: there are two chiral edge phases for $c_- = 16$, twenty-four chiral edge phases for $c_- =24$, more than one billion for $c_- =32$, and larger numbers of such edge phases for $c_- > 32$. The transition between the two possible chiral edge
phases was studied
in detail in the $c_- = 16$ case. \cite{Ginsparg87, Plamadeala13}

These fermionic and bosonic quantum Hall states illustrate the fact that the boundary-bulk correspondence
in topological states is not one-to-one. There can be multiple possible edge phases corresponding to the same bulk
phase. This can happen in a trivial way: two edge phases may differ by unstable gapless degrees
of freedom, so that one of the edge theories is more stable than the other.\cite{Haldane95,Kao99, Levin13, Lu13, Kane94}
(One interesting
refinement of this scenario is that the additional gapless degrees of freedom can be protected by a symmetry
so that, in the presence of this symmetry, both edge phases are stable\cite{Lu12}.) However, our focus
here is the situation in which there are multiple edge phases,
each of which is stable to weak perturbations without any symmetry considerations and none of which is more ``minimal" than the others.
In other words, in the integer and fractional quantum Hall states that we discuss here --
which have the additional property that they are all chiral -- all of the edge phases are
on the same footing. Although they can bound the same bulk, such edge phases generically
have different exponents and scaling functions for transport through point contacts and tunneling in from external leads. In some cases, the differences only show up in three-point and higher edge correlation functions.

In Sections \ref{sec:n=8}, \ref{sec:n=12} of this paper, we discuss fermionic integer quantum Hall states at
$\nu=8$ and $\nu=12$, their possible stable chiral edge phases, and the experimental signatures
that could distinguish these phases. In Section \ref{sec:fractional}, we discuss the simplest fractional quantum Hall states with multiple chiral edge phases, which occur at $\nu=8/7, 8/15, 16/5$ (fermions)
and $\nu=12/23$ (bosons).
Some of the edge phases that we construct do not
support gapless excitations with the quantum numbers of an electron. When the Hall conductance is non-zero,
the edge must have gapless excitations; in a system of electrons, there must be a finite-energy excitation
everywhere in the system with an electron's quantum numbers. However, it is not necessary that the electron be among the gapless edge excitations of an electronic quantum Hall state; it may be a gapped excitation at
the edge, above the gapless excitations that are responsible for carrying the Hall current.

Given the above statement that the same bulk phase can have multiple distinct chiral edge phases,
we should ask what breaks down in the usual relation between bulk topological phases and their associated edge spectra.
By the usual relationship, we mean the ``integration by parts" of a bulk Abelian Chern-Simons action that gives an edge
theory of chiral bosons with the same K-matrix \cite{Elitzur89,Wen92b}.
The answer is simply that the usual relation focuses only upon the lowest energy excitations of a system and ignores higher-energy excitations.
These higher-energy excitations are necessarily adiabatically connected to a topologically-trivial band insulator in the bulk
and, generically, gapped excitations at the edge.
Surprisingly, interactions between these ``trivial" modes and the degrees of freedom responsible for the topologically non-trivial state
can drive an edge phase transition that leads to a distinct edge phase without closing the bulk gap.
We refer to the relationship between these two distinct edge theories associated with the same bulk as {\it stable equivalence}.
At the level of the gapless edge modes, this manifests itself in the form of an edge reconstruction.
While the interpolation at the edge necessarily involves strong interactions, these can be understood using standard Luttinger liquid techniques.

The relationship between the edge and the bulk can also be viewed in the following manner. Each quasiparticle in the bulk has a topological
twist factor $\theta_a = e^{2\pi i h_a}$, with $0<{h_a}<1$. If the edge is fully chiral, each such quasiparticle corresponds to a tower of excitations.
The minimum scaling dimension for creating an excitation in this tower is $\text{min}\,\Delta_a = h_a + n_a$ for some integer $n_a$.
The other excitations in the tower are obtained by creating additional bosonic excitations on top of this minimal one; their scaling dimensions
are larger than the minimal one by integers. But if the edge has a different phase, the minimal scaling dimension operator in this tower
may be $\text{min}\,\Delta_a = h_a + {\tilde n}_a$. Therefore, the spectrum of edge operators can be different, even though the fractional parts
of their scaling dimensions must be the same. (In the case of a fermionic topological phase, we must compare scaling dimensions modulo $1/2$,
rather than modulo $1$. By fermionic topological phase, we mean one which can only occur in a system in which some of  the microscopic consitutents
are fermions. At a more formal level, this translates into the existence of a fermionic particle which braids trivially with all other particles.)

The purpose of this paper is to describe the precise conditions under which two different edge phases can terminate the same bulk state,
i.e. are stably equivalent.
These conditions are intuitive: the braiding statistics of the quasiparticle excitations of the bulk states must be identical
and the chiral central charges of the respective states must be equal.

Let us summarize the general relation between bulk Abelian topological states and their associated edge phases in slightly more mathematical terms.
Edge phases are described by lattices $\Lambda$ equipped with an integer-valued bilinear symmetric form $B$. \cite{Read90, frohlichzee91, Wen92a, frohlichstuderthiran, frohlichthiran94, frohlichmany95}
We collectively write this data as ${\cal E} = (\Lambda, B)$.
The signature of $B$ is simply the chiral central charge $c_-$ of the edge theory.
Given a basis $e_I$ for $\Lambda$, the bilinear form determines a K-matrix $K_{I J} = B(\vec{e_I}, \vec{e_J})$.
In a bosonic system, the lattice $\Lambda$ must be even while in a fermionic system, the lattice $\Lambda$ is odd.
(An odd lattice is one in which at least one basis vector has $(\text{length})^2$ equal to an odd integer.
The corresponding physical system will have a fermionic particle that braids trivially with all other particles.
This particle can be identified with an electron. An even lattice has no such vectors and, therefore, no fermionic particles
that braid trivially with all other particles. Hence, it can occur in a system in which none of the microscopic constituents are
fermions. Of course, a system, such as the toric code, may have fermionic quasiparticles that braid non-trivially with at least some other particles.)
Given the lattice $\Lambda$, vertex operators of the edge theory are associated with elements in the dual lattice $\Lambda^\ast$.
For integer quantum Hall states, $\Lambda^\ast = \Lambda$, however, for fractional states $\Lambda \subset \Lambda^\ast$.
The operator product expansion of vertex operators is simply given by addition in $\Lambda^\ast$.

Each bulk phase is characterized by the following data concisely written as ${\cal B} = (A, q, c_-\text { mod } 24)$:\cite{ frohlichzee91, frohlichstuderthiran, frohlichthiran94, frohlichmany95, Belov05,Stirling08} a finite Abelian group $A$ encoding the fusion rules for the distinct quasiparticle types, a finite quadratic form $q$ on $A$ that gives the topological spin to each particle type, and the chiral central charge modulo 24.
As we will discuss at length, since the map ${\cal E} \rightarrow {\cal B}$ associating edge data ${\cal E}$ to a given bulk ${\cal B}$ is {\it not} one-to-one, several different edge phases may correspond to the same bulk phase.
We will provide an in-depth mathematical description of the above formalism in order to precisely determine when two distinct edge phases correspond to the same bulk phase. To determine all of the edge phases that can bound the
same bulk, one can perform a brute force search through all lattices of a given dimension and determinant.
(For low-dimensional cases, the results of such enumeration is in tables in Ref.~\onlinecite{Conway88}
and in, for instance, G. Nebe's online Catalogue of Lattices.) Moreover, one can use a mass formula described
in Section \ref{sec:stable} to check if a list of edge phases is complete.

We will exemplify the many-to-one nature of the map ${\cal E} \rightarrow {\cal B}$ through various examples.
The most primitive example occurs for integer quantum Hall states.
For such states, the lattice is self-dual, $\Lambda^\ast = \Lambda$ so there are no non-trivial quasiparticles.
For $c_- < 8$, there is a unique edge theory for the fermionic integer quantum Hall state, however, at $c_- = 8$, there are two distinct lattices: the hypercubic latttice $\mathbb{I}_8$ and the $E_8$ root lattice.
Therefore, the associated gapless edge theories corresponding to each lattice may bound the same bulk state; there exists an edge reconstruction connecting the two edge phases.
Fractional states for which $A$ is non-trivial enrich this general structure.

A rather remarkable corollary of our analysis is the following: all rational Abelian topological phases in 2+1
dimensions can be described by Abelian Chern-Simons theory.
By rational, we mean that there is a finite number of bulk quasiparticle types, i.e., the group $A$ has finite order.
As may be seen by giving a physical interpretion to a theorem of Nikulin \cite{Nikulin80}
the particle types, fusion rules, and topological twist factors determine a genus of lattices,
from which we can define an Abelian Chern-Simons theory. A second result that follows from a theorem of Nikulin \cite{Nikulin80}
is that any fermionic Abelian topological phase can be mapped to a bosonic topological phase, together with some number of filled Landau levels.

The remainder of this paper is organized as follows.
We begin in Section \ref{sec:preliminaries} by reviewing the formalism used to describe the bulk and boundary excitations of Abelian Hall states.
As a means to both motivate the general mathematical structure and because of their intrinsic interest, we provide two examples of stable equivalence in the fractional quantum Hall setting in Section \ref{sec:illustrative} and summarize their physically distinct signatures.
In Section \ref{eqn:edge-transition}, we abstract from these two examples the general method for understanding how distinct edge phases of a single bulk are related via an edge phase transition.
In Section \ref{sec:stable}, we explain the bulk-edge correspondence through the concepts of stable equivalence and genera of lattices.
In Section \ref{sec:odd-even}, we explain how fermionic topological phases can be represented by bosonic topological phases together with some number of filled Landau levels.
In Section \ref{sec:examples}, we analyze observed integer and fractional quantum Hall states that admit multiple stable, fully chiral edge phases.
In Section \ref{sec:remarks}, we explain how a number of theorems due to Nikulin, that we use throughout the text, apply to the description of {\it all} Abelian topological field theories in (2+1)-D.
We conclude in Section \ref{sec:discussion}.
We have three appendices that collect ideas used within the text.

\section{Preliminaries}
\label{sec:preliminaries}

\subsection{Edge Theories}
\label{sec:edge-prelims}

In this section, we review the formalism that describes the edges of conventional
integer and Abelian fractional quantum Hall states.
We begin with the edges of fermionic integer quantum Hall states.
We assume that the bulks of these states are the conventional states that are
adiabatically connected to the corresponding states of non-interacting fermions.
As we will see in later sections, the edge structure is not uniquely determined,
even if we focus solely on chiral edge
phases that are stable against all weak perturbations.

All integer quantum Hall states
have one edge phase that is adiabatically connected to the edge of the corresponding
non-interacting fermionic integer quantum Hall state. This edge phase has effective action
${S_0} + {S_1}$, where
\begin{equation}
\label{eqn:fermionic}
{S_0} = \int dx dt \,{\psi_J^\dagger}\left(i\partial_t + A_t + {v_J}(i\partial_x + {A_x})\right){\psi_J}
\end{equation}
and $J = 1, 2, \ldots, N$.
We shall later study two interesting examples that occur when $N=8$ or $N = 12$. The operator $\psi^\dagger_J$
creates an electron at the edge in the $J^{\rm th}$ Landau level;
$v_J$ is the edge velocity of an electron in the $J^{\rm th}$ Landau level.
Inter-edge interactions take the form
\begin{multline}
\label{eqn:fermionic-inter-edge}
{S_1} = \int dx \,dt \, \bigl(t_{JK}(x) \,e^{i \left({k_F^J}-{k_F^K}\right)x}\,\psi^\dagger_J \psi_K + \text{h.c.}\\ +
v_{JK}{\psi_J^\dagger}{\psi_J} {\psi_K^\dagger}{\psi_K} + \ldots \bigr).
\end{multline}
The $\ldots$ in Eq.~(\ref{eqn:fermionic-inter-edge}) represent higher-order tunneling and interaction
terms that are irrelevant by power counting. We neglect these terms and focus on the first two
terms. Electrons in different Landau levels will generically have different Fermi momenta.
When this is the case, the tunneling term (the first term in Eq. (\ref{eqn:fermionic-inter-edge}))
will average to zero in a translationally-invariant system.
In the presence of disorder, however, $t_{IJ}(x)$ will be random and relevant (e.g. in a replicated action
which is averaged over $t_{IJ}(x)$). Moreover, it is possible for the Fermi momenta to be equal;
for instance, in an $N$-layer system in which each layer has a single filled Landau level,
the Fermi momenta will be the same if the electron density is the same in each layer.
Fortunately, we can make the change of variables:
$$
{\psi_J}(x) \rightarrow \left(\overline{\cal P}\exp\left(i\int_{-\infty}^x dx' M(x')\right)\right)_{JK}{\psi_K}(x),
$$
where $M(x)$ is the matrix with entries $M_{JK}=t_{JK}(x') \,e^{i \left({k_F^J}-{k_F^K}\right)x'}/\overline{v}$,
$\overline{v}=\sum_J {v_J}/N$, and $\overline{\cal P}$ denotes anti-path-ordering. When this is substituted into Eq.~(\ref{eqn:fermionic}), the first term in Eq.~(\ref{eqn:fermionic-inter-edge}) is eliminated from the action
$S_0 + S_1$.
This is essentially a U(N) gauge transformation that gauges away inter-mode scattering. An extra random kinetic
term proportional to $(v_J-\overline{v})\delta_{IJ}$ is generated, but this is irrelevant in the infrared when disorder-averaged.

The second term in Eq.~(\ref{eqn:fermionic-inter-edge}) is an inter-edge density-density interaction;
$v_{JK}$ is the interaction between edge electrons in the $J^{\rm th}$ and $K^{\rm th}$ Landau levels.
This interaction term can be solved by bosonization.
The action $S_0 + S_1$ from Eqs. (\ref{eqn:fermionic}) and (\ref{eqn:fermionic-inter-edge}) can be equivalently represented by the bosonic action
\begin{multline}
\label{eqn:bosonic}
S =  \int dx\,dt \biggl(\frac{1}{4\pi}\delta_{IJ} \partial_t \phi^I \partial_x \phi^J - \frac{1}{4\pi}V_{IJ}\partial_x \phi^I \partial_x \phi^J\\ +
\frac{1}{2\pi} \sum_I \epsilon_{\mu\nu} \partial_\mu \phi^I A_\nu
\biggr),
\end{multline}
where $V_{II}\equiv {v_I} + v_{II}$ (no summation) and $V_{IJ}\equiv v_{IJ}$ for $I\neq J$.
The electron annihilation operator is bosonized according to $\psi_J \sim \eta_J e^{i \phi^J}$.
Here $\eta_J$ is a ``Klein factor'' satisfying $\eta_J \eta_K = - \eta_K \eta_J$ for $J\neq K$, which ensures
that $\psi_J \psi_K = - \psi_K \psi_J$. Products of even numbers of Klein factors can be diagonalized
and set to one of their eigenvalues, $\pm 1$, if all terms in the Hamiltonian commute with them.
They can then be safely ignored. This is the case in all of the models studied in this paper.
This action can be brought into the following diagonal form (setting the external electromagnetic field
to zero for simplicity):
\begin{equation}
S = \int dx\,dt \left(\frac{1}{4\pi}\delta_{IJ} \partial_t {\tilde \phi}^I \partial_x {\tilde \phi}^J -
\frac{1}{4\pi}{v_I}\delta_{IJ}\partial_x {\tilde \phi}^I \partial_x {\tilde \phi}^J
\right)
\end{equation}
with an orthogonal transformation $\phi^I = O^{I{ }}_J {\tilde \phi}^J$ that diagonalizes
$V_{IJ}$ according to $O^{I{ }}_L V_{IJ} O^{J{ }}_K = {\tilde v_L}\delta_{LK}$.
Two-point correlation functions take the form
\begin{equation}
\label{vertex-op-corr-fcn}
\left\langle e^{i m_I \phi^I} e^{-i m_K \phi^K}\right\rangle
= \prod_{J=1}^{N}\frac{1}{(x-{\tilde v_J}t)^{m_I m_K O^{I{ }}_J  O^{K{ }}_J}}.
\end{equation}
There is no sum over $J$ in the exponent on the right-hand-side of Eq.~(\ref{vertex-op-corr-fcn}).
The electron Green function in the $I^{\rm th}$ Landau level is a special case of this
with ${m_K} = \delta_{IK}$.


It is now straightforward to generalize the preceding discussion to the case of
an arbitrary Abelian integer or fractional quantum Hall state \cite{Wen92b}. For simplicity, we will
focus on the case of fully chiral phases in which all edge modes move in the same direction.
Such phases do not, in general, have a free fermion representation and can only be described
by a chiral Luttinger liquid. They are characterized by equivalence classes of
positive-definite symmetric integer \emph{K-matrices} $K$, and integer \emph{charge vectors} $t$ that enter the chiral Luttinger liquid action
according to
\begin{multline}
\label{eqn:bosonic-general}
S_{LL} = \int dx\, dt \biggl(\frac{1}{4\pi}K_{IJ} \partial_t \phi^I \partial_x \phi^J -
\frac{1}{4\pi}V_{IJ}\partial_x \phi^I \partial_x \phi^J\\ + \frac{1}{2\pi} t_I \epsilon_{\mu\nu} \partial_\mu \phi^I A_\nu
\biggr).
\end{multline}
The fields in this action satisfy the periodicity condition $\phi^I \equiv \phi^I + 2\pi n^I$
for $n^I \in \mathbb{Z}$.
Two phases, characterized by the pairs $({K_1},{t_1})$ and $({K_2},{t_2})$,
are equivalent if $K_1 = W^T K_2 W$ and ${t_1}={t_2}W$, where
$W\in \GL(N,\mathbb{Z})$ since the first and third terms in the two theories
can be transformed into each other by the change of variables $\phi^I = W^I_{\ J}{\tilde \phi}^J$.
So long as $W\in \GL(N,\mathbb{Z})$, the periodicity condition satisfied by ${\tilde \phi}^J$
is precisely the same as the periodicity condition satisfied by $\phi^I$.
The matrix $V_{IJ}$ consists of marginal deformations that do not change the phase
of the edge but affect the propagation velocities. (If we wish, we can think of each phase
as a fixed surface under RG flow, and the $V_{IJ}$s are marginal deformations that
parametrize the fixed surface.) All such chiral edge theories are stable to
all weak perturbations by the same reasoning by which we analyzed integer quantum Hall edges.
The simplest fermionic fractional quantum Hall edge theory is that of the
Laughlin $\nu=1/3$ state, for which $K = (3)$ and $t=(1)$ (a $1\times 1$ matrix
and a $1$-component vector, respectively). Integer quantum Hall edges are the special
case, $K_{IJ}=\delta_{IJ}$ or, allowing for basis changes, $K=W^T W$ with $W \in \GL(N, \mathbb{Z})$.

It is useful to characterize these phases by lattices $\Lambda$ rather than equivalence classes
of K-matrices. Let $e_{I}^a$ be the eigenvector of $K$ corresponding to eigenvalue
$\lambda_a$: $K_{IJ} e_{J}^a = \lambda^a e_{I}^a$. We normalize $e_{J}^a$ so that
$e_{J}^a e_{J}^b = \delta^{ab}$ and define a metric $g_{ab} = \lambda_a \delta_{ab}$.
Then, $K_{IJ} =  g_{ab} e_{I}^a e_{J}^b$ or, using vector notation, $K_{IJ} =  {\bf e}_I \cdot {\bf e}_J$.
We will be focusing mostly on positive-definite lattices, so that $g_{ab}$ has signature $(N,0)$
but we will occasionally deal with Lorentzian lattices, for which we take $g_{ab}$ has signature $(p, N-p)$.
The metric $g_{a b}$ defines a {\it bilinear form} $B$ on the lattice $\Lambda$ (and its dual $\Lambda^\ast$) --
this just means we can multiply two lattice vectors $\vec{e}_I, \vec{e}_J$ together using the metric, $\vec{e}_I \cdot \vec{e}_J = e_I^a g_{a b} e_J^b = B(\vec{e}_I, \vec{e}_J)$.
The $N$ vectors ${\bf e}_I$ define a lattice $\Lambda = \{ m_I {\bf e}_I  | {m_I} \in \mathbb{Z}\}$.
The $\GL(N,\mathbb{Z})$ transformations $K \rightarrow W^T K W$ are simply basis changes
of this lattice, so we can equally well describe edge phases by equivalence classes of $K$-matrices
or by lattices $\Lambda$. The conventional edge phases of integer quantum Hall states described
above correspond to hypercubic lattices $\mathbb{Z}^N$, which we will often denote by
the corresponding $K$ matrix in its canonical basis, $\mathbb{I}_N$.
The $\nu=1/3$ Laughlin state corresponds to the lattice
$\Lambda= \mathbb{Z}$ with dual $\Lambda^\ast = \frac{1}{3}\mathbb{Z}$.
\footnote{This statement assumes the periodicity convention, $\phi \equiv \phi + 2 \pi n$, for $n \in \mathbb{Z}$.}
The connection of quantum Hall edge phases to lattices
can be exploited more easily if we make the following change of variables,
$X^a = e_{I}^a \phi^I$, in terms of which the action takes the form
\begin{multline}
\label{eqn:bosonic-X1}
S = \frac{1}{4\pi} \int dx\,dt \biggl(g_{ab} \partial_t X^a \partial_x X^b -
v_{ab}\partial_x X^a \partial_x X^b.
\biggr)
\end{multline}
The variables $X^a$ satisfy the periodicity condition ${\bf X} \equiv {\bf X} + 2\pi {\bf y}$
for ${\bf y} \in \Lambda$ and $v_{ab}\equiv V_{IJ} f^I_{a} f^J_{b}$,
where $f_{a}^I$ are basis vectors for the dual lattice $\Lambda^*$, satisfying
$f^I_{a} e_J^a = e_{L a} (K^{-1})^{L I} e_J^a = \delta^I_J$.


Different edge phases (which may correspond to different bulks or the same bulk; the latter is
the focus of this paper) are distinguished by their correlation functions.
The periodicity conditions on the fields $X^a$ dictate that the allowed exponential
operators are of the form $e^{i {\bf v}\cdot{\bf X}}$, where ${\bf v}\in \Lambda^*$.
These operators have scaling dimensions
\begin{equation}
\text{dim}\!\left[ e^{i {\bf v}\cdot{\bf X}} \right] = \frac{1}{2}|{\bf v}|^2.
\end{equation}
They obey the operator algebra
\begin{eqnarray}
:e^{i {\bf v_1} \cdot {\bf X}}: :e^{i {\bf v_2} \cdot {\bf X}}: \sim :e^{i ({\bf v_1} + {\bf v_2}) \cdot {\bf X}}:,
\end{eqnarray}
where $:\cdot:$ denotes normal ordering.
Thus, the operator spectrum and algebra is entirely determined by the underlying dual lattice $\Lambda^\ast$.

In a quantum Hall state, there are two complementary ways of measuring some of the
scaling exponents. The first is a quantum point contact (QPC)
at which two edges of a quantum Hall fluid are brought together at a point so that
quasiparticles can tunnel across the bulk from one edge to the other. Even though a single
edge is completely stable against all weak perturbations, a pair of oppositely-directed
edges will, in general, be coupled by relevant perturbations
\begin{equation}
S = S_T + S_B + \int dt \, \sum_{{\bf v}\in \Lambda^*}
v_{\bf v} \,e^{i {\bf v}\cdot\left[{\bf X}_T -{\bf X}_B \right]}.
\end{equation}
Here, $T,B$ are the two edges, e.g., the top and bottom edges of a Hall bar; we will use this notation throughout whenever it is necessary to distinguish the two edges.
The renormalization group (RG) equation for $v_{\bf v}$ is
\begin{equation}
\frac{d v_{\bf v}}{d\ell} = \left(1-{\left|{\bf v}\right|^2}\right)v_{\bf v}.
\end{equation}
If ${\bf v}\cdot {\bf f}^I t_I\neq 0$, the above coupling transfers ${\bf v}\cdot {\bf f}^I t_I$ units of charge across the junction and this perturbation will contribute
to the backscattered current according to
\begin{equation}\label{eq:backscattering}
I^b \propto \left|v_{\bf v}\right|^2\,V^{2{\left|{\bf v}\right|^2} - 1}.
\end{equation}
A second probe is the tunneling current from a metallic lead:
\begin{eqnarray*}
S &=& S_{\rm edge} + S_{\rm lead} \\
& &\hspace{.4in} +\, \int dt \sum_{{\bf v}\in \Lambda}
t_{\bf v} \Bigl[\psi_{\rm lead}^\dagger \partial\psi_{\rm lead}^\dagger
{\partial^2}\psi_{\rm lead}^\dagger \ldots\Bigr] e^{i {\bf v}\cdot {\bf X}}.
\end{eqnarray*}
The term in square brackets $[ ... ]$ contains $n$ factors of $\psi_{\rm lead}^\dagger$ and $n(n-1)/2$ derivatives,
where $n={\bf v}\cdot {\bf f}^I t_I$ must be an integer.
The RG equation for $t_{\bf v}$
\begin{equation}
\frac{d t_{\bf v}}{d\ell} = \left(1-\frac{n^2}{2}-\frac{1}{2}{\left|{\bf v}\right|^2}\right)t_{\bf v}.
\end{equation}
The contribution to the tunneling current from $t_{\bf v}$ (assuming $n\neq 0$) is
\begin{equation}\label{eq:tun-current-polarized}
I^{\rm tun} \propto \left|t_{\bf v}\right|^2\,V^{{\left|{\bf v}\right|^2}+{n^2} - 1}.
\end{equation}
Here, we have assumed that the spins at the edge of the quantum Hall state are fully spin-polarized
and that tunneling from the lead conserves $S_z$. If, however, either of these conditions is violated,
then other terms are possible in the action. For instance, charge-$2e$ tunneling can take the form
\begin{equation}
t_{\rm pair} \int dt \,\psi_{{\rm lead},\uparrow}^\dagger \psi_{{\rm lead},\downarrow}^\dagger
\, \,  e^{i {\bf v}\cdot {\bf X}},
\end{equation}
where ${\bf v}\cdot {\bf f}^I t_I = 2$. Then, we have tunneling current
\begin{equation}\label{eq:tun-current-unpolarized}
I^{\rm tun} \propto \left|t_{\bf v}\right|^2\,V^{{\left|{\bf v}\right|^2} + 1}.
\end{equation}

Generically, two lattices $\Lambda_1$ and $\Lambda_2$ can be distinguished by
the possible squared lengths  $|{\bf v}|^2$ for ${\bf v}\in \Lambda_1^*$.
In many cases of interest, the shortest length, which will dominate
the backscattered current discussed above, is enough to distinguish
two edge phases of the same bulk.
However, sometimes, as in the case of the two bosonic
integer quantum Hall states with $c=16$ discussed in Ref.~\onlinecite{Plamadeala13}
the spectrum of operator scaling dimensions (not just the shortest length, but
all lengths along with degeneracies at each length level) is precisely the same in the two theories,
so they could only be distinguished by comparing three-point correlation functions.
In either case, different edge phases can be distinguished by their correlation functions.


\subsection{Bulk Theories}
\label{sec:bulk-prelims}

In a later section, we will explain how bulk phases correspond to the mathematical notion of a {\it genus} of lattices, while their associated edge theories are given by lattices within a genus (or in the case of fermionic theories, a pair of genera, one odd and one even).
In order to explain the relation between the genus of a lattice
and a bulk Abelian phase, we recall some facts
about Abelian topological phases.

Suppose that we have a $2+1$d Abelian topological phase
associated to a lattice $\Lambda$.
Choosing a basis  ${{\bf e}_I}$ for the lattice $\Lambda$, we define
$K_{IJ} = {{\bf e}_I}\cdot{{\bf e}_J}$ and write
a bulk effective action
\begin{eqnarray}
\label{cstheory}
{\cal S} = \int d^3x \Big({1 \over 4 \pi} \epsilon^{\mu \nu \rho} K_{I J} a^I_\mu \partial_\nu a^J_\rho
+ {1 \over 2 \pi} j^{\mu}_I a_\mu^I \Big).
\end{eqnarray}
A particle in this theory carrying charge
$m_I$ under the gauge field $a_I$ can be associated with a vector ${\bf v} \equiv m_I {\bf f}^I$,
where ${\bf f}_I$ is the basis vector of $\Lambda^*$ dual to ${\bf e}_I$ and satisfying $(K^{-1})^{IJ} {\bf e}_J = {\bf f}^I$.
Recall that because $\Lambda \subset \Lambda^\ast$, any element in $\Lambda$ can be expressed in terms of the basis for $\Lambda^\ast$, however, the converse is only true for integer Hall states for which $\Lambda = \Lambda^\ast$.
Particles ${\bf v}$, ${\bf v'} \in \Lambda^*$
satisfy the fusion rule ${\bf v}\times {\bf v'} = {\bf v} + {\bf v'}$ and their braiding results in the multiplication of the wave function describing the state by an overall phase
$
e^{2\pi i {\bf v}\cdot{\bf v'}}$.
Since this phase is invariant under shifts ${\bf v}\rightarrow {\bf v} + \boldsymbol{\lambda}$ for $\boldsymbol{\lambda}\in \Lambda$,
the topologically-distinct particles are associated with
elements of the so-called discriminant group $A={\Lambda^*}/\Lambda$.
The many-to-one nature of the edge-bulk
correspondence is a reflection of the many-to-one correspondence between
lattices $\Lambda$ and their discriminant groups $A$.
Equivalent bulk phases necessarily have identical discriminant groups so our initial choice of lattice is merely a representative in an equivalence class of bulk theories.

We now define a few terms. A {\it bilinear symmetric form} on a finite Abelian group $A$ is a function $b \colon A \times A \to \bQ/\bZ$ such that for every $a,a',a'' \in A$, 
\[b(a+a',a'') = b(a,a'') + b(a',a'')\] and $b(a,a') = b(a',a)$.  As all bilinear forms considered in this paper will be symmetric, we will simply call them \emph{bilinear forms} with symmetric being understood.  A \emph{quadratic form} $q$ on a finite Abelian group $A$ is a function $q: A \rightarrow \mathbb{Q}/\mathbb{Z}$ such that  $q(na)=n^2q(a)$ for every $n\in \mathbb{Z}$, and such that 
\[q(a+a') - q(a) - q(a') = b(a,a')\] for some bilinear form $b\colon A \times A \to \bQ/\bZ$.  In this case, we say that $q$ \emph{refines} $b$, or is a \emph{quadratic refinement} of $b$.  A bilinear $b$ or quadratic form $q$ is \emph{degenerate} if there exists a non-trivial subgroup $S \subset A$ such that $b(s,s') = 0$ or $q(s) = 0$ for every $s,s' \in S$.  Throughout this paper, all bilinear and quadratic forms will be assumed nondegnerate.
Each K-matrix $K$ determines a symmetric bilinear form $B$ on $\bR^n$ via $B({\bf x},{\bf y}) = {\bf x}^T K {\bf y}$ that takes integer values on the lattice $\bZ^n \subset \bR^n$.  Every other lattice $\Lambda \subset \bR^n$ on which $B$ is integral can be obtained by acting on $\bZ^n$ by the orthogonal group $\{g \in \GL(N,\bR) : g K g^T = K\}$ of $K$.  On the other hand, an integral symmetric bilinear form is equivalent to a lattice according to the construction before Eq. \eqref{eqn:bosonic-X1} in Section \ref{sec:edge-prelims}.  We are therefore justified in  using the terminology ``lattice'' and ``K-matrix'' in place of ``integral symmetric bilinear form'' throughout this paper.  Every diagonal entry of a K-matrix $K$ is even iff the $(\text{length})^2$ of every element in the lattice $\bZ^N$ is even.  We call $K$ \emph{even} if this is the case, and otherwise it is \emph{odd}.  Even K-matrices determine integral quadratic forms on $\bZ^N$ via $Q({\bf x}) = \frac 12 {\bf x}^T K {\bf x}$, while for odd K-matrices they are half-integral.  When we simply write bilinear or quadratic form or, sometimes, \emph{finite bilinear form} or \emph{finite quadratic form}, we will mean a nondegenerate symmetric bilinear form, or nondegenerate quadratic form, whose domain is a finite Abelian group.  Throughout, we abbreviate the ring $\bZ/N\bZ$ of integers modulo $N$ as $\bZ/N$.

The $S$-matrix of the theory can be given in terms of the elements of the discriminant group:
\begin{equation}
S_{{[\bf v]},{[\bf v']}} =\frac{1}{\sqrt{|A|}} e^{- 2\pi i {\bf v}\cdot{\bf v'}} = {1 \over \sqrt{|A|}} e^{- 2 \pi i m_I (K^{-1})^{IJ} m'_J},
\end{equation}
where ${\bf v} = m_I {\bf f}^I,{\bf v'} = m'_J {\bf f}^J \in \Lambda^*$ and $|A|$ is the dimension of the discriminant group.
The bracketed notation $[\vec{v}]$ indicates an equivalence class of elements $[\vec{v}] \in \Lambda^\ast/\Lambda = A$.
Our normalization convention is to represent elements in the dual lattice $\Lambda^\ast$ with integer vectors $m_I$.
The bilinear form $B$ on $\Lambda^*$ reduces modulo $\Lambda$ to define a finite bilinear form on the discriminant group $\Lambda^\ast/\Lambda$ via
\[b([m_I {\bf f}^I], [m'_J {\bf f}^J]) = B(m_I {\bf f}^I, m'_J {\bf f}^J) = m_I (K^{-1})^{IJ} m'_J.\]
The topological twists $\theta_{[\bf v]}$, which are the eigenvalues of the $T$ matrix,
are defined by
\begin{equation}
\label{eqn:T-c-theta}
T_{{[\bf v]},{[\bf v']}} = e^{-\frac{2\pi i}{24}{c_-}}\,\theta_{[\bf v]}\, \delta_{{[\bf v]},{[\bf v']}}
\end{equation}
where
\begin{equation}
\theta_{[\bf v]} =  e^{\pi i {\bf v}\cdot{\bf v}}.
\end{equation}
Note that Eq. (\ref{eqn:T-c-theta}) implies that the theory is invariant under shifts of $c_-$ by $24$ so long as the topological twists
$\theta_{[\bf v]}$ are invariant, but its modular transformation properties, which determine the partition function on $3$-manifolds
via surgery \cite{Witten89}, is sensitive to shifts by $c_- \neq 0$ (mod $24$).

If the topological twists are well-defined on the set of quasiparticles $A$,
then they must be invariant under ${\bf v}\mapsto{\bf v} + \boldsymbol{\lambda}$, where
$\boldsymbol{\lambda}\in \Lambda$, under which
\begin{equation}
\theta_{[{\bf v}]} \mapsto \theta_{[{\bf v + \lambda}]} = \theta_{[{\bf v}]} \,e^{\pi i \boldsymbol{\lambda}\cdot\boldsymbol{\lambda}}.
\end{equation}
If the K-matrix is even, so that we are dealing with a bosonic theory, $\boldsymbol{\lambda}\cdot\boldsymbol{\lambda}$ is even for all
$\boldsymbol{\lambda}\in \Lambda$. If the K-matrix is odd, however -- i.e. if the
system is fermionic -- then there are some $\boldsymbol{\lambda}\in \Lambda$ for which
$\boldsymbol{\lambda}\cdot\boldsymbol{\lambda}$ is odd.
In this case, the topological twists are not
quite well-defined, and more care must be taken, as we describe in Section \ref{sec:odd-even}.
Given the above definition, only $T^2$ is well-defined.

In a bosonic Abelian topological phase, we can define a finite quadratic form $q$
on the discriminant group, usually called the {\it discriminant form}, according to
\begin{equation}
  q([\vec{v}])=\frac{1}{2}\vec{v}^2 = {1 \over 2} m_I (K^{-1})^{IJ} m_J \text{ mod } \bZ,
  \label{eqn:quadratic-form-twists}
\end{equation}
where $\vec{v} = m_I {\bf f}^I$.
In a topological phase of fermions, we will have to define $q$ with more care,
as we discuss in Section \ref{sec:odd-even}.
Thus, we postpone its definition until then and will only discuss Abelian bosonic topological phases in the remainder of this section.
In terms of the discriminant form $q$, the $T$-matrix takes the form
\begin{equation}
\label{eqn:T-from-q}
 \theta_{a}  = e^{2\pi i q(a)},
\end{equation}
and the $S$-matrix takes the form
\begin{eqnarray}
S_{a,a'} &=& \frac{1}{\sqrt{|A|}} e^{2\pi i (q(a-a') - q(a) - q(-a'))} \\ 
 &=& \frac{1}{\sqrt{|A|}} e^{-2\pi i (q(a+a') - q(a) - q(a'))} \label{eqn:S-from-q}
\end{eqnarray}
The equation for the $S$-matrix makes use of the fact that the finite bilinear form $b$ can be recovered from
the finite quadratic form according to $b(a,a') = q(a+a') - q(a) - q(a')$.
(It is satisfying to observe that the relation between the bilinear form $b$ and the discriminant form $q$ coincides exactly with the phase obtained by a wave function when two particles are twisted about one another.)
While the introduction of the discriminant form may appear perverse in the bosonic context, we will find it to be an essential ingredient when discussing fermionic topological phases.

In any bosonic topological phase, the chiral central charge is related to the
bulk topological twists by the following relation \cite{KitaevHoneycomb}:
\begin{equation}
  \frac{1}{\mathcal{D}}\sum_a d_a^2\theta_a=e^{2\pi ic_-/8}.
  \label{}
\end{equation}
Here $\mathcal{D}=\sqrt{\sum_a d_a^2}$ is the total quantum dimension, $d_a$ is the quantum dimension of the quasiparticle type $a$, and $\theta_a$ is the corresponding topological twist/spin. $c_- =c-\ol{c}$ is the chiral central charge.
In an Abelian bosonic phase described by an even matrix $K$, the formula simplifies to
\begin{equation}
  \frac{1}{\sqrt{|A|}}\sum_{a\in A}e^{2\pi iq(a)}=e^{2\pi i c_-/8},
  \label{eqn:Gauss-Milgram}
\end{equation}
since $d_a = 1$ for all quasiparticle types.
Here $|A|=\sqrt{|\det K|}$ and $c_- = {r_+}-{r_-}$ is the signature of the matrix,
the difference between the number of positive and negative eigenvalues.
(We will sometimes, as we have done here, use the term signature to refer to the difference ${r_+}-{r_-}$,
rather than the pair $({r_+},{r_-})$; the meaning will be clear from context.)
Notice that $e^{2\pi iq(a)}$ is just the topological twist of the quasiparticle represented by $a \in \Lambda^*/\Lambda$. This is known as the Gauss-Milgram sum in the theory of integral lattices.

Let us pause momentarily to illustrate these definitions in a simple example: namely, the semion theory described by the K-matrix, $K = (2)$.
This theory has discriminant group $A = \mathbb{Z}/2 \mathbb{Z} = Z_2$ and, therefore, two particle types, the vacuum denoted by the lattice vector $[0]$ and the semion $s = [1]$.
Recall that our normalization convention is to take the bilinear form on $A$ to be $b([\vec{x}], [\vec{y}]) = x \cdot {1 \over 2} \cdot y$;
the associated quadratic form is then $q([\vec{x}]) = {1 \over 2} b([\vec{x}], [\vec{x}])$.
The discriminant form, evaluated on the semion particle, is given by $q([1]) = {1 \over 2} \cdot {1^2 \over 2}$.
The T matrix equals $\exp(-2\pi i /24){\rm diag}(1, i)$, and the S-matrix, $S = \frac{1}{\sqrt{2}}
\smat{1 & 1 \\ 1 & -1}$.  Evaluating the Gauss-Milgram sum confirms that $c_- = 1$.

In order to determine the discriminant group from a given $K$-matrix,
we can use the following procedure.
First, we compute the Gauss-Smith normal form of the $K$-matrix, which can be found using a standard algorithm\cite{Cohen}.
Given $K$, this algorithm produces
integer matrices $P$, $Q$, $D$ such that
\begin{equation}
K=PDQ.
\end{equation}
Here both $P$ and $Q$ are unimodular $|\!\det P|=|\!\det Q|=1$, and $D$ is diagonal.  The diagonal entries of $D$ give the orders of a minimal cyclic decomposition of the discriminant group
\[A \simeq \prod_J \bZ/D_{JJ},\]
with the fewest possible cyclic factors, giving yet another set of generators for the quasiparticles.  Although more compact, this form does not directly lend itself towards checking the equivalence of discriminant forms.

Now recall that the bases of $\Lambda$ and $\Lambda^*$ are related by $K$:
\begin{equation}
{\bf e}_I = K_{IJ} {\bf f}^J
\end{equation}
Substituting the Gauss-Smith normal form, this can be rewritten
\begin{equation}
(P^{-1})^{IL} {\bf e}_L = D^{IK} Q_{KJ} {\bf f}^J.
\end{equation}
The left-hand side is just a basis change of the original lattice.
On the right-hand side, the row vectors of $Q$ that correspond to entries of $D$ greater than $1$ give the generators of the cyclic subgroups of the discriminant group. A non-trivial example is given in Appendix \ref{sec:finding-discriminant}.

\section{Two Illustrative Examples of Bulk Topological Phases with Two Distinct Edge Phases}
\label{sec:illustrative}

The chiral Luttinger liquid action is stable against all small perturbations involving only the gapless
fields in the action in Eq.~(\ref{eqn:bosonic-general}) (or, equivalently in the integer case,
the action in Eq.~(\ref{eqn:fermionic})).
This essentially follows from the chirality of the theory,
but it is instructive to see how this plays out explicitly.\cite{Wengapless}
However, this does not mean that a given bulk will have only a single edge phase.\cite{chamonwenreconstruction}
A quantum Hall system will have additional gapped excitations which we can ignore only if
the interactions between them and the gapless excitations in Eq.~(\ref{eqn:bosonic-general})
are weak.
If they are not weak, however, we cannot ignore them and interactions with these degrees of freedom can lead to an edge phase transition \cite{Plamadeala13}.

We will generally describe the gapped excitations with a K-matrix equal to $\sigma_z = \smat{1 & 0 \cr 0 & -1}$.
We may imagine this K-matrix arising from a thin strip of $\nu = 1$ fluid living around the perimeter of our starting Hall state.\cite{chamonwenreconstruction}
For edge phase transitions between bosonic edges theories, we should instead take the gapped modes to be described by a K-matrix equal to $\sigma_x = \smat{0 & 1 \cr 1 & 0}$.
It is important to realize that the existence of the localized (gapped) edge modes described by either of these K-matrices implies the appropriate modification to the Chern-Simons theory describing the bulk topological order.
This addition does not affect the bulk topological order\cite{WittenJones}; without symmetry, such a gapped state is adiabatically connected to a trivial band insulator.

We will illustrate this with two concrete examples.
We begin with the general edge action
\begin{multline}
S = \int dx\, dt \biggl(\frac{1}{4\pi}K_{IJ} \partial_t \phi^I \partial_x \phi^J \\ -
\frac{1}{4\pi}V_{IJ}\partial_x \phi^I \partial_x \phi^J + \frac{1}{2\pi} t_I \epsilon_{\mu\nu} \partial_\mu \phi^I A_\nu
\biggr).
\end{multline}
The first example is described by the K-matrix
\begin{equation}
\label{eqn:twelve-elevenths}
  K_1 =
  \begin{pmatrix}
	1 & 0 \\
	0 & 11
  \end{pmatrix},
\end{equation}
with $t=(1, -1)^T$. This is not an example that is particularly
relevant to quantum Hall states observed in experiments -- we will discuss several examples
of those in Section \ref{sec:examples} -- but it is simple and serves as a paradigm for the more
general structure that we discuss in Sections \ref{sec:stable} and \ref{sec:odd-even}.

Let us suppose that we have an additional left-moving and additional right-moving fermion which, together, form a gapped unprotected excitation.
The action now takes the form
\begin{multline}
\label{eqn:fractional-stable-equiv}
S = \int dx\, dt \biggl(\frac{1}{4\pi}\left({K_1} \oplus {\sigma_z}\right)_{IJ}
\partial_t \phi^I \partial_x \phi^J \\ -
\frac{1}{4\pi}V_{IJ}\partial_x \phi^I \partial_x \phi^J + \frac{1}{2\pi} t_I \epsilon_{\mu\nu} \partial_\mu \phi^I A_\nu
\biggr),
\end{multline}
where we have now extended $t=(1, -1, 1, 1)^T$.
The K-matrix for the two additional modes is taken to be $\sigma_z$. 
We will comment on the relation to the
$\sigma_x$ case in Sections \ref{eqn:edge-transition} and \ref{sec:stable}.

If the matrix $V_{IJ}$ is such that the perturbation
\begin{equation}
S' = \int dx \, dt \, u' \cos({\phi_3} + {\phi_4}),
\end{equation}
is relevant, and if this is the only perturbation added to Eq.~(\ref{eqn:fractional-stable-equiv}),
then the two additional modes become gapped and the system is in the phase (\ref{eqn:twelve-elevenths}).
Suppose, instead, that the only perturbation is
\begin{equation}
S'' = \int dx \, dt \, u'' \cos({\phi_1}-11{\phi_2}+2{\phi_3}+4{\phi_4}).
\end{equation}
This perturbation is charge-conserving and spin-zero (i.e., its left and right scaling dimensions are equal).
If it is relevant, then the edge is in a different phase. To find this phase, it is helpful to make
the basis change:
\begin{equation}
W^T (K_1\oplus\sigma_z) W=K_2\oplus\sigma_z,
\end{equation}
where
\begin{equation}
\label{eqn:twelve-elevenths-alt}
K_2 =
\left(
\begin{array}{cc}
 3 & 1  \\
 1 & 4  \\
\end{array}
\right),
\end{equation}
and
\begin{equation}
W=
\left(
\begin{array}{cccc}
 0 & 0 & 1 & 0 \\
 0 & -2 & 0 & 1 \\
 -2 & 3 & 0 & -2 \\
 1 & -7 & 0 & 4 \\
\end{array}
\right).
\end{equation}
Making the basis change $\phi=W\phi'$, we see that
\begin{equation}
{\phi_1}-11{\phi_2}+2{\phi_3}+4{\phi_4} = \phi_3' + \phi_4'.
\end{equation}
Therefore, the resulting phase is described by (\ref{eqn:twelve-elevenths-alt}).

To see that these are, indeed, different phases, we can compute basis-independent quantities,
such as the lowest scaling dimension of any operator in the two theories. In the $K_1$ theory,
it is $1/22$ while in the $K_2$ theory, it is $3/22$.
Measurements that probe the edge structure in
detail can, thereby, distinguish these two phases of the edge. Consider, first, transport through a
QPC that allows tunneling between the two edges of the Hall bar, as described in Sec~\ref{sec:edge-prelims}.
In the state governed by $K_1$, the most relevant backscattering term is $\cos(\phi_2^T -\phi_2^B)$. Applying Eq~(\ref{eq:backscattering}), the backscattered current will depend on the voltage according to
\begin{equation}
I^b_{1} \propto V^{-9/11}.
\end{equation}
An alternative probe is given by tunneling into the edge from a metallic lead. The most relevant term in the $K_1$ edge phase that tunnels one electron into the lead is $\psi^\dagger_{\rm lead}e^{i\phi_1^T} $. Applying Eq~(\ref{eq:tun-current-polarized}) yields the familiar current-voltage relation,
\begin{equation}
I_1^{\rm tun} \propto V.
\end{equation}

In contrast, in the phase governed by $K_2$, the most relevant backscattering term across a QPC is given by $\cos(\phi_2'^T - \phi_2'^B)$, which from Eq~(\ref{eq:backscattering}) yields the current-voltage relation
\begin{equation}
I^b_{2} \propto V^{-5/11},
\end{equation}
while the most relevant single-electron tunneling term is given by $\psi_{\rm lead}^\dagger e^{-3i\phi_1'^T-i\phi_2'^T}$, which yields the scaling from Eq~(\ref{eq:tun-current-polarized})
\begin{equation}
I_2^{\rm tun} \propto V^{3}.
\end{equation}

Since the two edge theories given by $K_1$ and $K_2$ are connected by a phase transition just on the edge, we may expect they bound the same bulk Chern-Simons theory. Indeed, the bulk quasiparticles can be identified up to ambiguous signs due to their fermionic nature. First, the discriminant group of the $K_1$ theory is $\mathbb{Z}/{11}$. We define a quasiparticle basis for this theory as $\psi_{j}\equiv (-j, - 6j)^T, j=0, 1, \dots, 10$.
(While the cyclic nature of the group $\mathbb{Z}/{11}$ implies the identification $(a, b) \equiv (a', b')\ {\rm mod}\ (1,11)$ for $a, b, a', b' \in \mathbb{Z}$, we choose the above basis in order to ensure charge conservation.)
The $S$ matrix is given by $S_{jj'}=\frac{1}{\sqrt{11}}e^{- \frac{72\pi i}{11}jj'}$. For the other theory given by $K_2$, the discriminant group obviously has the same structure with the generator being $(0, 1)^T$ and the quasiparticles are denoted by $\psi_j'$. The $S$ matrix is given by $S'_{jj'}=\frac{1}{\sqrt{11}}e^{- \frac{6\pi i}{11}jj'}$. Now we make the following identification:
\begin{equation}
  \psi_j'\longleftrightarrow \psi_{j}.
  \label{}
\end{equation}
This identification preserves the $\U(1)$ charge carried by each quasiparticle.
The $S$ matrices are also identified:
\begin{equation}
  S_{j,j'}=\frac{1}{\sqrt{11}}e^{- \frac{72\pi i}{11}jj'}=\frac{1}{\sqrt{11}}e^{- \frac{6\pi i}{11}jj'}=S'_{jj'}.
  \label{}
\end{equation}
Since the diagonal elements of $S$ are basically $T^2$, it follows that the topological spins are also identified up to $\pm 1$.

Our second example is
\begin{equation}
\label{eqn:eight-sevenths}
  K'_1=
  \begin{pmatrix}
	1 & 0 \\
	0 & 7
  \end{pmatrix},
\end{equation}
with $t=(1,1)^T$.
As before, we suppose that a non-chiral pair of modes comes down in energy and interacts strongly
with the two right-moving modes described by (\ref{eqn:eight-sevenths}). The action now takes the form
\begin{multline}
\label{eqn:fractional-stable-equiv2}
S =  \int dx\, dt \biggl(\frac{1}{4\pi}\left({K'_1} \oplus {\sigma_z}\right)_{IJ}
\partial_t \phi^I \partial_x \phi^J \\ 
- \frac{1}{4\pi}V_{IJ}\partial_x \phi^I \partial_x \phi^J + \frac{1}{2\pi} t_I \epsilon_{\mu\nu} \partial_\mu \phi^I A_\nu
\biggr).
\end{multline}
If the matrix $V_{IJ}$ is such that the perturbation
\begin{equation}
S' = \int dx \, dt \, u' \cos({\phi_3}+{\phi_4})
\end{equation}
is relevant and this is the only perturbation added to Eq.~(\ref{eqn:fractional-stable-equiv2}),
then the two additional modes become gapped and the system is in the phase in Eq. (\ref{eqn:eight-sevenths}).
Suppose, instead, the only perturbation is the following:
\begin{equation}
S'' = \int dx \, dt \, u'' \cos({\phi_1}+7{\phi_2}+{\phi_3}+3{\phi_4}).
\end{equation}
This perturbation is charge-conserving and spin-zero.
If it is relevant, then the edge is in a different phase. To find this phase, it is helpful to make
the basis change
\begin{equation}
W'^T (K'_1\oplus\sigma_z) W'=K'_2\oplus\sigma_z,
\end{equation}
where
\begin{equation}
\label{eqn:eight-sevenths-alt}
  K'_2=
  \begin{pmatrix}
	2 & 1\\
	1 & 4
  \end{pmatrix}
\end{equation}
and
\begin{equation}
W'=
\begin{pmatrix}
	2 & 1 & 0 & -1\\
	1 & -1 & 0 & -1\\
	0 & 0 & -1 & 0\\
	-3 & 2 & 0 & 3
  \end{pmatrix}.
\end{equation}
Making the basis change $\phi=W'\phi'$, we see that
\begin{equation}
\phi_1+7\phi_2+\phi_3+3\phi_4 =\phi_4'-\phi_3'.
\end{equation}
Therefore, the resulting phase is described by (\ref{eqn:eight-sevenths-alt}).
This is a different phase, as may be seen by noting that
the lattice corresponding to Eq.~(\ref{eqn:eight-sevenths-alt})
is an even lattice while the lattice corresponding to Eq.~(\ref{eqn:eight-sevenths}) is odd.

The difference between the two edge phases is even more dramatic than in
the previous example.
One edge phase has gapless fermionic excitations while the other one does not!
This example shows that an edge reconstruction can relate a theory with fermionic topological order to one with bosonic topological order.
Again, these two edge phases of the $\nu=8/7$ can be distinguished by the voltage dependence of the
current backscattered at a quantum point contact and the tunneling current from a metallic lead.
In the $K_1'$ edge phase (\ref{eqn:eight-sevenths}), the backscattered current at a QPC is dominated by the tunneling term $\cos(\phi_2^T - \phi_2^B)$; using Eq~(\ref{eq:backscattering}) this yields the current-voltage relation
\begin{equation}
I_1^b \propto V^{-5/7},
\end{equation}
while the single-electron tunneling into a metallic lead is dominated by the tunneling term $\psi_{\rm lead}^\dagger e^{i\phi_1^T}$, which, using Eq~(\ref{eq:tun-current-polarized}), yields the familiar linear current-voltage scaling
\begin{equation}
I_1^{\rm tun} \propto V.
\end{equation}
In the $K_2'$ edge phase (\ref{eqn:eight-sevenths-alt}), the backscattered current at a QPC is dominated by the backscattering term $\cos(\phi_2'^T - \phi_2'^B)$, yielding:
\begin{equation}
I_2^b \propto V^{-3/7}.
\end{equation}
The tunneling current from a metallic lead is due to the tunneling of charge-$2e$ objects
created by the edge operator $e^{i\phi'_1 + 4 i \phi'_2}$. If we assume that the electrons are fully spin-polarized
and $S_z$ is conserved, then the most relevant term that tunnels $2e$ into the metallic lead is $\psi_{\rm lead}^\dagger\partial\psi_{\rm lead}^\dagger e^{i\phi'^T_1 + 4 i \phi'^T_2}$. Using Eq~(\ref{eq:tun-current-polarized}) the tunneling current is proportional to a very high power of the voltage:
\begin{equation}
I_2^{\rm tun} \propto V^{7}.
\end{equation}

Again, although the theories look drastically different, we can show that the bulk $S$ matrices are isomorphic.  First, the discriminant group of the $K_1'$ theory is $\mathbb{Z}/{7}$ whose generator we can take to be the $(0, 4)$ quasiparticle. We label all quasiparticles in this theory as $\psi_j \equiv (0, 4 j), j=0, 1, \dots, 6$. The $S$ matrix is given by $S_{jj'}=\frac{1}{\sqrt{7}}e^{- \frac{32\pi i}{7}jj'}$. For the other theory given by $K_2'$, the discriminant group is generated by $(0, 1)^T$ and we denote the quasiparticles by $\psi_j'$. The $S$ matrix is given by $S'_{jj'}=\frac{1}{\sqrt{7}}e^{- \frac{4\pi i}{7}jj'}$. Now we make the following identification:
\begin{equation}
  \psi_j'\longleftrightarrow \psi_{j}.
  \label{}
\end{equation}
 The $S$ matrices are then seen to be identical:
\begin{equation}
  S_{j,j'}=\frac{1}{\sqrt{7}}e^{- \frac{32\pi i}{7}jj'}=\frac{1}{\sqrt{7}}e^{- \frac{4\pi i}{7}jj'}=S'_{jj'}.
  \label{}
\end{equation}

\section{Edge Phase Transitions}
\label{eqn:edge-transition}

In the previous section, we gave two simple examples of edge phase transitions that can occur between two distinct chiral theories.
In this section, we discuss how edge transitions can occur in full generality.

The chiral Luttinger liquid action is stable against all perturbations involving only the gapless
fields in the action in Eq.~(\ref{eqn:bosonic-general}) (or, equivalently in the integer case,
the action in Eq.~(\ref{eqn:fermionic})). However, as we have seen in the previous section,
strong interactions with gapped excitations can drive a phase transition that occurs
purely at the edge. While the bulk is completely unaffected, the edge undergoes a
transition into another phase.

On the way to understanding this in more generality,
we first consider an integer quantum Hall state. At the edge of such
a state, we expect additional gapped excitations that we ordinarily ignore.
However, they can interact with gapless excitations.
(Under some circumstances, they can even become gapless.\cite{chamonwenreconstruction})
Let us suppose that we have an additional left-moving and
and additional right-moving fermion which, together, form a gapped unprotected excitation.
Then additional terms must be considered in the action.
Let us first consider the case of an integer quantum Hall edge. The action in
Eqs.~(\ref{eqn:fermionic}) and (\ref{eqn:fermionic-inter-edge}) becomes $S_{0}+S_{1}+S_{u}$ with
\begin{multline}
\label{eqn:fermionic-plus-gapped}
S_{u} = \int dx \,dt \Bigl(\psi_{N+1}^\dagger \left(i\partial_t + v_{N+1}i\partial_x\right) \psi^{}_{N+1}\\
+ \psi_{N+2}^\dagger \left(i\partial_t - v_{N+2}i\partial_x\right) \psi^{}_{N+2}\\
+ u \psi_{N+1}^\dagger \psi^{}_{N+2} + \text{h.c.}\\
+ v_{I,N+1} {\psi_I^\dagger}{\psi^{}_I} {\psi_{N+1}^\dagger}{\psi^{}_{N+1}}
+ v_{I,N+2} {\psi_I^\dagger}{\psi^{}_I} {\psi_{N+2}^\dagger}{\psi^{}_{N+2}}\\
+ {\cal L}_{N,L} \Bigr),
\end{multline}
where $\psi_{N+1}$, $\psi_{N+2}$ annihilate right- and left-moving excitations which
have an energy gap $u$ for $v_{I,N+1}=v_{I,N+2}=0$. So long as $v_{I,N+1}$ and $v_{I,N+2}$
are small, this energy gap survives, and we can integrate out $\psi_{N+1}$, $\psi_{N+2}$, thereby
recovering the action ${S_0}+{S_1}$ in Eqs.~(\ref{eqn:fermionic}) and (\ref{eqn:fermionic-inter-edge}),
but with the couplings renormalized. However, if $v_{I,N+1}$ and $v_{I,N+2}$ are sufficiently large,
then some of the other terms in the action, which we have denoted by ${\cal L}_{N,L}$ in
Eq.~(\ref{eqn:fermionic-plus-gapped}) may become more relevant than $u$.
These include terms such as
\begin{equation}
{\cal L}_{N,L} = {u_I}\psi_{I}^\dagger \psi^{}_{N+2} + \text{h.c.} + \ldots.
\end{equation}

In order to understand these terms better, it is helpful to switch to the bosonic representation,
where there is no additional overhead involved in considering the general case of
a chiral Abelian state, integer or fractional:
\begin{multline}
\label{eqn:bosonic-N+2}
S = \int dx\,dt \biggl(\frac{1}{4\pi}\left({K \oplus \sigma_z}\right)^{}_{IJ} \partial_t \phi^I \partial_x \phi^J - 
\frac{1}{4\pi}V_{IJ}\partial_x \phi^I \partial_x \phi^J\\ +
\sum_{m_I} u_{m_I} \cos\left({m_I}{\phi^I}\right)
+ \frac{1}{2\pi} \sum_I \epsilon_{\mu\nu} \partial_\mu \phi^I A_\nu
\biggr).
\end{multline}
Here, $I=1,2,\ldots,N+2$; and $\left({K \oplus \sigma_z}\right)^{}_{IJ}$
is the direct sum of $K$ and $\sigma_z$:
$\left({K \oplus \sigma_z}\right)^{}_{IJ}=K_{IJ}$ for $I=J=1,2,\ldots,N$,
$\left({K \oplus \sigma_z}\right)^{}_{IJ}=1$ for $I=J=N+1$,
$\left({K \oplus \sigma_z}\right)^{}_{IJ}=-1$ for $I=J=N+2$,
and $\left({K \oplus \sigma_z}\right)^{}_{IJ}=0$ if $I\in\{1,2,\ldots,N\}$, $J\in\{N+1,N+2\}$
or vice-versa. The interaction matrix has $V_{I,N+1}\equiv v_{I,N+1}$, $V_{I,N+2}\equiv v_{I,N+2}$.
The $m^I$s must be integers because the $\phi^I$s are periodic.
For instance, ${m_I}=(0,0,\ldots,0,1,-1)$ corresponds to the mass term $u (\psi_{N+1}^\dagger \psi^{}_{N+2} + \text{h.c.})$ in Eq.~(\ref{eqn:fermionic-plus-gapped}),
so $u_{m_I}=u$. In the last term, we are coupling
all modes equally to the electromagnetic field, i.e. this term can be written
in the form $t_I \epsilon_{\mu\nu} \partial_\mu \phi^I A_\nu$ with $t_I =1$ for all $I$.
This is the natural choice, since we expect additional fermionic excitations to carry electrical
charge $e$.

In general, most of the couplings $u_{m_I}$ will be irrelevant at the Gaussian fixed point.
An irrelevant coupling cannot open a gap if it is small enough to remain in the basin
of attraction of the Gaussian fixed point.
However, if we make the coupling large enough, it may be in the basin of attraction of another
fixed point and it may open a gap. We will not comment more on this possibility here. However, we can imagine tuning the $V_{IJ}$s so that any given $u_{m_I}$ is relevant. To analyze this possibility, it is helpful to change to the variables
$X^a = e_{I}^a \phi^I$, in terms of which the action takes the form
\begin{multline}
\label{eqn:bosonic-X2}
S =  \int dx\,dt \biggl(\frac{1}{4\pi}\eta_{ab} \partial_t X^a \partial_x X^b -
\frac{1}{4\pi}v_{ab}\partial_x X^a \partial_x X^b\\ +
\sum_{m_I} u_{m_I} \cos\left({m_I} {f^I_a} X^a\right)
+ \frac{1}{2\pi} \sum_I {f^J_a} \epsilon_{\mu\nu} \partial_\mu  X^a A_\nu.
\biggr)
\end{multline}
$e_I^a$ and $f^I_a$ are bases for the lattice $\Lambda_{N+2}$ and its dual $\Lambda_{N+2}^\ast$, where the lattice $\Lambda_{N+2}$ corresponds to $K \oplus \sigma_z $.
The variables $X^a$ satisfy the periodicity condition ${\bf X} \equiv {\bf X} + 2\pi {\bf y}$
for ${\bf y} \in \Lambda_{N+2}$. Note that, since one of the modes is left-moving,
the Lorentzian metric $\eta_{ab} = {\rm diag}({\bf 1}_{N-1}, - 1)$ appears in Eq. (\ref{eqn:bosonic-X2}).

Since ${f^I_a}$ is a basis of the dual lattice $\Lambda_{N+2}^*$,
the cosine term can also be written in the form
\begin{equation*}
\sum_{{\bf v}\in \Lambda_{N+2}^*} u_{\bf v} \cos\left( {\bf v} \cdot {\bf X}\right).
\end{equation*}
The velocity/interaction matrix is given by $v_{ab}=V_{IJ}{f^I_a}{f^J_b}$.
Now suppose that the velocity/interaction matrix takes the form
\begin{equation}
\label{eqn:V-choice}
v_{ab} =  v\,O^c_{\ a} \delta_{cd} O^d_{\ b},
\end{equation}
where $O\in \SO(N+1,1)$. Then we can make a change of variables
to ${\tilde X}^a \equiv O^a_{\ b} X^b$.
We specialize to the case of a single cosine perturbation associated with
a particular vector in the dual lattice ${\bf v}_0 \equiv {p_I} {\bf f}^I$ which we
will make relevant (we have also set $A_\nu =0$ since it is inessential
to the present discussion). Now Eq.~(\ref{eqn:bosonic-X2}) takes the form
\begin{multline}
S = \frac{1}{4\pi} \int dx\,dt \biggl(\eta_{ab} \partial_t \tilde{X}^a \partial_x \tilde{X}^b -
v \delta_{ab}\partial_x \tilde{X}^a \partial_x \tilde{X}^b\\ +
u_{{\bf v}_0} \cos\left({p_I} {f^I_a} (O^{-1})^a_{\ b} \tilde{X}^b\right)
\biggr).
\end{multline}
If this perturbation has equal right and left scaling dimensions (i.e., is spin-zero), then its scaling dimension is simply
twice its left scaling dimension with corresponding beta function
\begin{equation}
\frac{du_{{\bf v}_0}}{d\ell} = \left(2-q_{N+2}^2 \right) u_{{\bf v}_0},
\end{equation}
where $q_b \equiv {p_I} {f^I_a} (O^{-1})^a_{\ b}$. The transformation $O^{-1}$ can be chosen to
be a particular boost in the $(N+2)$-dimensional space $\mathbb{R}^{N+1,1}$.
Because $q_a$ is a null vector (i.e., a light-like vector) in this space, by taking the boost in the opposite direction of the
``spatial" components of $q_a$, we can ``Lorentz contract'' them, thereby making $q_{N+2}$
as small as desired. Thus, by taking $v_{ab}$ of the form (\ref{eqn:V-choice}) and
choosing $O\in \SO(N+1,1)$ so that $q_{N+2}^2 <2$, we can make this coupling relevant.

When this occurs, two modes, one right-moving and one left-moving,
will acquire a gap. We will then be left over with a theory with $N$ gapless right-moving modes.
The gapless excitations $\exp(i{\bf v}\cdot {\bf X})$ of the system must commute with
${{\bf v}_0}\cdot {\bf X}$ and, since the cosine fixes ${{\bf v}_0}\cdot {\bf X}$,
any two excitations that differ by ${{\bf v}_0}\cdot {\bf X}$ should be identified. Thus, the resulting
low-energy theory will be associated with the lattice $\Gamma$
defined by $\Gamma \equiv {\Lambda_\perp}/\Lambda_\parallel$, where
${\Lambda_\perp}, {\Lambda_\parallel} \subset \Lambda_{N+2}$ are defined by
${\Lambda_\perp} \equiv \{{\bf v}\in\Lambda_{N+2}\, |\, {\bf v}\cdot {\bf v}_0 = 0 \}$
and ${\Lambda_\parallel} \equiv \{n{{\bf v}_0}\, |\, n\in \mathbb{Z}\}$.
If ${\bf g}_I$ is a basis for $\Gamma$, then we can define a K-matrix
in this basis, ${\tilde K}_{IJ}={{\bf g}_I}\cdot{\bf g}_J$. The low-energy effective
theory for the gapless modes is
\begin{multline}
\label{eqn:bosonic-resulting}
S =  \int dx\, dt \biggl(\frac{1}{4\pi}{\tilde K}_{IJ} \partial_t \phi^I \partial_x \phi^J -
\frac{1}{4\pi}\tilde{V}_{IJ}\partial_x \phi^I \partial_x \phi^J\\ + \frac{1}{2\pi}\tilde{t}_I \epsilon_{\mu\nu} \partial_\mu \phi^I A_\nu
\biggr).
\end{multline}
When ${\bf v}_0 = (0,0,\ldots,0,1,-1)$ is the only relevant operator,
$\phi^{N+1}$ and $\phi^{N+2}$ are gapped out. Therefore, $\Gamma=\Lambda$ and
${\tilde K}_{IJ}=K_{IJ}$. However, when other operators are present,
$\Gamma$ could be a different lattice $\Gamma\ncong\Lambda$,
from which it follows that ${\tilde K}_{IJ}\neq K_{IJ}$ (and, ${\tilde K}\neq W^T K W$
for any $W$).

We motivated the enlargement of the theory from $K$ to $K\oplus\sigma_z$
by assuming that an additional pair of gapped counter-propagating
fermionic modes comes down in energy and interacts strongly with the gapless
edge excitations. This counter-propagating pair of modes can be viewed as a thin
strip of $\nu=1$ integer quantum Hall fluid or, simply, as a fermionic Luttinger liquid.
Of course, more than one such pair of modes may interact strongly with the gapless
edge excitations, so we should also consider enlarging the K-matrix to
$K\oplus{\sigma_z}\oplus{\sigma_z}\ldots\oplus{\sigma_z}$.
We can generalize this by imagining that we can add
any one-dimensional system to the edge of a quantum Hall state. (This may not be
experimentally-relevant to presently observed quantum Hall states, but as a matter of principle,
this is something that could be done without affecting the bulk, so
we should allow ourselves this freedom.) Any clean, gapless 1D system of fermions
is in a Luttinger liquid phase (possibly with some degrees of freedom gapped).
Therefore, $K\oplus{\sigma_z}\oplus{\sigma_z}\ldots\oplus{\sigma_z}$
is actually the most general possible form for the edge theory.

One might wonder about the possibility of attaching a thin strip of a fractional quantum Hall state
to the edge of the system. Naively, this would seem to be a generalization of our
putative most general form $K\oplus{\sigma_z}\oplus{\sigma_z}\ldots\oplus{\sigma_z}$.
To illustrate the issue, let us consider a bulk $\nu=1$ IQH state and place a thin strip
of $\nu=1/9$ FQH state at its edge. The two edges that are in close proximity can be
described by the following K-matrix:
\begin{equation}
K=
\begin{pmatrix}
1 & 0\\
0 & -9
\end{pmatrix}.
\end{equation}
As discussed in Ref. \onlinecite{Levin13}, this edge theory can become fully gapped with charge-non-conserving backscattering. Then we are left with the outer chiral edge of the thin strip, which
is described by $K=(9)$, which can only bound a topologically ordered $\nu=1/9$ Laughlin state.
The subtlety here is that a thin strip of the fractional quantum Hall state has no two-dimensional bulk and should be considered as a purely one-dimensional system. Fractionalized excitations, characterized by fractional conformal spins only make sense when a true 2D bulk exists.
If the width of the strip is small,
so that there is no well-defined bulk between them, then we can only allow operators that add an integer
number of electrons to the two edges. We cannot add fractional charge since there is no bulk which
can absorb compensating charge. Thus the minimal conformal spin of any operator
is $1/2$. In other words, starting from an one-dimensional interacting electronic system, one cannot change the conformal spin of the electron operators. So attaching a thin strip of FQH state is no different from attaching a trivial pair of modes.

In a bosonic system, we cannot even enlarge our theory by a pair of counter-propagating fermionic
modes. We can only enlarge our theory by a Luttinger
liquid of bosons or, equivalently, a thin strip of $\sigma_{xy}=\frac{2e^2}{h}$ bosonic integer quantum hall fluid \cite{Lu12,Levin13, Senthil13}.
Such a system has K-matrix equal to $\sigma_x$, which only has bosonic excitations. Equivalently,
bosonic systems must have even $K$-matrices -- matrices with only even
numbers along the diagonal -- because all particles that braid trivially with every other
particle must be a boson. Since the enlarged matrix must have the same determinant as the original
one because the determinant is the ground state degeneracy of the bulk phase on the
torus \cite{Wen92a}, we can only enlarge the theory by $\sigma_x$, the minimal even unimodular matrix.
Therefore, in the bosonic case, we must enlarge our theory by $K\rightarrow K\oplus\sigma_x$.

In the fermionic case, we must allow such an enlargement by $\sigma_x$ as well. We can imagine the fermions forming pairs and these pairs forming a bosonic Luttinger liquid which enlarges $K$
by $\sigma_x$. In fact, it is redundant to consider both $\sigma_z$ and $\sigma_x$: for an odd matrix $K$, $W(K\oplus\sigma_z)W^T = K\oplus\sigma_x$,
where
\begin{equation}
  W=
  \begin{pmatrix}
	 1 & 0 & \cdots & 0  & y_1 & -y_1\\
	 0 & 1 & \cdots & 0 &  y_2 & -y_2\\
	 \vdots & \vdots & \vdots & \vdots & \vdots & \vdots\\
	 0 & 0 & \cdots & 1 & y_N & -y_N\\
	 0 & 0 & \cdots  & 0 & 1 & -1 \\
	 x_1 & x_2 & \cdots & x_N & s & 1-s \\
  \end{pmatrix}
  \label{}
\end{equation}
Here the vector $\vec{x}$ has an odd length squared, i.e. $\vec{x}^TK\vec{x}$ is odd; by definition of $K$ odd, such an $\vec{x}$ must exist. The vector $\vec{y}$ is defined as $\vec{y}=-K\vec{x}$ and the integer $s$ by $s=\frac{1}{2}(1-\vec{x}^TK\vec{x})$.
Thus $K\oplus\sigma_x$ is $\GL(N+2,\mathbb{Z})$-equivalent to $K\oplus\sigma_z$ and our previous discussion for fermionic systems could be redone entirely with extra modes
described by $\sigma_x$. However, if $K$ is even,
then $K\oplus\sigma_x$ is not $\GL(N+2,\mathbb{Z})$-equivalent to $K\oplus\sigma_z$.

We remark that although $\sigma_z$ enlargement and $\sigma_x$ enlargement are equivalent for
fermionic states when topological properties are concerned, they do make a difference in charge vectors: the appropriate charge vector for the $\sigma_z$ block should be odd and typically taken to be $(1,1)^T$. However the charge vector for the $\sigma_x$ block must be even and needs to be determined from the similarity transformation.

To summarize, a quantum Hall edge phase described by matrix $K_1$ can undergo a purely
edge phase transition to another edge phase with $\GL(N,\mathbb{Z})$-inequivalent
$K_2$ (with identical bulk) if there exists
${\tilde W}\in \GL(N+2k,\mathbb{Z})$
such that
\begin{equation}
{K_2}\oplus\sigma_x\oplus\ldots\oplus\sigma_x =
{\tilde W}^T \left({K_1}\oplus{\sigma_x}\oplus\ldots\oplus\sigma_x \right){\tilde W}.
\end{equation}
for some number $k$ of $\sigma_x$s on each side of the equation.
In a fermionic system with $K_1$ odd, an edge phase transition can also occur
to an even matrix $K_2$ if
\begin{equation}
\label{eqn:sigma-z-stable}
{K_2^{\rm even}}\oplus\sigma_z\oplus\ldots\oplus\sigma_x =
{\tilde W}^T \left({K_1^{\rm odd}}\oplus{\sigma_x}\oplus\ldots\oplus\sigma_x \right){\tilde W}.
\end{equation}


\section{Stable Equivalence, Genera of Lattices, and the Bulk-Edge Correspondence
for Abelian Topological Phases}
\label{sec:stable}

\subsection{Stable Equivalence and Genera of Lattices}

In the previous section, we saw that a bulk Abelian quantum Hall state
associated with $K_1$ has more than one different stable chiral
edge phase if there exists $\GL(N,\mathbb{Z})$-inequivalent
$K_2$ and ${\tilde W}\in \GL(N+2k,\mathbb{Z})$ such that
\begin{equation}
\label{eqn:sigma-x-stable}
{K_2}\oplus\sigma_x\oplus\ldots\oplus\sigma_x =
{\tilde W}^T \left({K_1}\oplus{\sigma_x}\oplus\ldots\oplus\sigma_x \right){\tilde W}.
\end{equation}
This is an example of a stable equivalence; we say that $K_1$ and $K_2$ are \emph{stably equivalent} if,
for some $n$, there exist signature $(n,n)$ unimodular matrices $L_i$
such that $K_1 \oplus L_1$ and $K_2 \oplus L_2$ are \emph{integrally equivalent},
i.e. are $\GL(N+2n,\mathbb{Z})$-equivalent.
If there is a choice of $L_i$s such that both are even,
we will say that $K_1$ and $K_2$ are ``$\sigma_x$-stably equivalent''
since the $L_i$s can be written as direct sums of $\sigma_x$s.
We also saw in Eq.~\ref{eqn:sigma-z-stable} that when $K_1$ is
odd and $K_2$ is even, we will need $L_2$ to be an odd matrix.
We will call this ``$\sigma_z$-stable equivalence" since $L_2$ must contain
a $\sigma_z$ block.
We will use $U$ to denote the signature $(1,1)$ even Lorentzian lattice associated with $\sigma_x$.
Then $\sigma_x$-stable equivalence can be restated in the language of lattices as follows.
Two lattices $\Lambda_1$, $\Lambda_2$ are $\sigma_x$-stably equivalent if
${\Lambda_1}\oplus U \cdots \oplus U$, and $\Lambda_2 \oplus U \cdots \oplus U$
are isomorphic lattices. Similarly, $U_z$ will denote the Lorentzian lattice associated with $\sigma_z$.
Occasionally, we will abuse notation and use $\sigma_x$ and $\sigma_z$ to refer to the corresponding lattices $U$, $U_z$.

Stable equivalence means that the two
$K$-matrices are equivalent after adding ``trivial'' degrees of freedom -- i.e. purely 1D
degrees of freedom that do not require any change to the bulk. This is analogous to the notion
of stable equivalence of vector bundles, according to which two vector bundles are stably equivalent if
and only if isomorphic bundles are obtained upon joining them with trivial bundles.

We now introduce the concept of the \emph{genus} of a lattice or integral quadratic form.
Two integral quadratic forms are in the same \emph{genus}\cite{Cassels78,Conway88} when they have the same signature and are equivalent over the \emph{$p$-adic integers} $\bZ_p$ for every prime $p$.  Loosely speaking, equivalence over $\bZ_p$ can be thought of as equivalence modulo arbitrarily high powers of $p$, i.e.\ in $\bZ/p^n$ for every $n$.  The importance of genus in
the present context stems from the following statement of Conway and Sloane \cite{Conway88}:
\medskip 

\noindent 
{\it Two integral quadratic forms $K_1$ and $K_2$
are in the same genus if and only if ${K_1}\oplus\sigma_x$ and ${K_2}\oplus\sigma_x$
are integrally equivalent. }
\medskip 

Proofs of this statement are, however, difficult to pin down in the literature.  It follows, for instance, from results in Ref. \onlinecite{Cassels78} about a refinement of the genus called the spinor genus.   Below, we show how it follows in the even case from results stated by Nikulin\cite{Nikulin80}.
This characterization of the genus is nearly the same as the definition of $\sigma_x$-stable equivalence
given in (\ref{eqn:sigma-x-stable}), except that Eq. (\ref{eqn:sigma-x-stable}) allows multiple copies
which is natural since a physical system may have access to multiple copies of trivial degrees of freedom.
Its relevance to our situation follows from the following theorem that we demonstrate below:
\medskip

\noindent
{\it Two $K$-matrices $K_1$ and $K_2$ of the same dimension, signature and type are stably
equivalent if and only if ${K_1}\oplus\sigma_x$ and ${K_2}\oplus\sigma_x$
are integrally equivalent, i.e.\ only a single copy of $\sigma_x$ is needed in Eq. (\ref{eqn:sigma-x-stable}).}
\medskip

Thus any edge phase that can be reached via a phase transition involving multiple sets of
trivial 1D bosonic degrees of freedom (described by K-matrix $\sigma_x$) can also be reached
through a phase transition involving only a single such set. We demonstrate this by appealing to the following
result stated by Nikulin\cite{Nikulin80} (which we paraphrase but identify by his numbering):
\medskip

\noindent
Corollary 1.16.3: {\it The genus of a lattice is determined by its discriminant group $A$,
parity, signature $(r_+,r_-)$, and bilinear form $b$ on the discriminant group.}
\medskip

Since taking the direct sum with multiple copies of $\sigma_x$ does not change the parity, or bilinear form on the discriminant group, any $K_1$ and $K_2$ that are $\sigma_x$-stably equivalent
are in the same genus. The theorem then follows from the statement\cite{Conway88} above that only a single copy of $\sigma_x$ is needed. 

In the even case, the theorem follows directly  from two other results found in  Nikulin\cite{Nikulin80}:
\medskip

\noindent
Corollary 1.13.4: {\it For any even lattice $\Lambda$ with signature $({r_+},{r_-})$ and discriminant quadratic
form $q$, the lattice $\Lambda\oplus U$ is the only lattice with
signature $({r_+}+1,{r_-}+1)$ and quadratic form $q$.}
\smallskip

\noindent
Theorem 1.11.3: {\it Two quadratic forms on the discriminant group are isomorphic
if and only if their bilinear forms are isomorphic and they have the same signature} (mod 8).
\medskip

If lattices $\Lambda_1$ and $\Lambda_2$ are in the same genus, they must have the same
$({r_+},{r_-})$ and bilinear form $b$. According to Theorem 1.11.3, they must have the same
quadratic form, namely $q([{\bf x}])=\frac{1}{2}b([{\bf x}],[{\bf x}])$, which is well-defined in the case
of an even lattice. Then, Corollary 1.13.4 tells us that ${\Lambda_1}\oplus U$ is the unique lattice
with signature $({r_+}+1,{r_-}+1)$ and quadratic form $q$. Since ${\Lambda_2}\oplus U$ has the
same signature $({r_+}+1,{r_-}+1)$ and quadratic form $q$, ${\Lambda_1}\oplus U\cong
{\Lambda_2}\oplus U$. Thus, we see that any two even $K$-matrices in the same
genus are integrally-equivalent after taking the direct sum with a single copy of $\sigma_x$.
Of course, our previous arguments that used Nikulin's Corollary~1.16.3
and the characterization of genus from Conway and Sloane\cite{Conway88}
are stronger since they apply to odd matrices.

\subsection{Bulk-Edge Correspondence}

Since the quadratic form $q([\vec{u}])$ gives the $T$ and $S$ matrices
according to Eqs. (\ref{eqn:T-from-q}) and (\ref{eqn:S-from-q}),
we can equally-well say that the genus of a
lattice is completely determined by the particle types, $T$-matrix, $S$-matrix, and right- and left-central charges.
For a bosonic system, the genus completely determines a bulk phase.
Conversely, a bulk topological phase almost completely determines a genus: the bulk phase
determines $(c_+ - c_-)$ mod 24 while a genus is specified by $({c_+},{c_-})$.
However, if the topological phase is fully chiral, so that it can have $c_- = 0$,
then it fully specifies a family of genera that differ only by adding central charges that
are a multiple of $24$, i.e. $3k$ copies of the $E_8$ state for some integer $k$ (see Section \ref{sec:n=8}
for a discussion of this state). Thus, up to innocuous shifts of the central charge by $24$, we can say that
\medskip

\noindent
{\it A bulk bosonic topological phase corresponds to a genus of even lattices while its edge phases
correspond to the different lattices in this genus.}
\medskip

\noindent
The problem of detemining the different stable edge phases that can occur for the same bosonic bulk
is then the problem of determining how many distinct lattices there are in a genus.

In the fermionic case, the situation is more complicated. A fermionic topological phase is determined
by its particle types, its $S$-matrix, and its central charge (mod 24). It does not have a well-defined $T$-matrix
because we can always change the topological twist factor of a particle by $-1$ simply by adding an electron to it.
According to the following result of Nikulin, these quantities determine an odd lattice:
\medskip

\noindent
Corollary 1.16.6: {\it Given a finite Abelian group $A$, a bilinear form $b: A\times A \rightarrow \mathbb{Q}/\mathbb{Z}$,
and two positive numbers $({r_+},{r_-})$, then, for sufficiently large ${r_+},{r_-}$, there exists an odd lattice for which $A$ is its discriminant group.
$b$ is the bilinear form on the discriminant group, and $({r_+},{r_-})$ is its signature.}
\medskip

Since the $S$-matrix defines a bilinear form on the Abelian group of particle types, this theorem means that the quantities that specify
a fermionic Abelian topological phase are compatible with an odd lattice. Clearly, they are also compatible with an entire genus of odd lattices
since $\sigma_x$ stable equivalence preserves these quantities. Moreover, by Corollary 1.16.3, there is only a single genus of odd lattices
that are compatible with this bulk fermionic Abelian topological phase. However, Corollary 1.16.3 leaves open the possibility that
there is also a genus of even lattices that is compatible with this fermionic bulk phase, a possibility that was realized in one of the examples
in Section \ref{sec:illustrative}. This possibility is discussed in detail in Section \ref{sec:odd-even}. However, the general result that we can
already state, up to shifts of the central charge by $24$ is

\medskip
\noindent
{\it A bulk fermionic topological phase corresponds to a genus of odd lattices while its edge phases
correspond to the different lattices in this genus and, in some cases (specificed in Section \ref{sec:odd-even}),
to the different lattices in an associated genus of even lattices.}
\medskip

In principle, one can determine how many lattices there are in a given genus
by using the Smith-Siegel-Minkowski mass formula \cite{Conway88} to evaluate the weighted sum
\begin{equation}
  \sum_{\Lambda\in g} \frac{1}{|\text{Aut}(\Lambda)|} = m(K)
  \label{eqn:massformula1}
\end{equation}
over the equivalence classes of lattices in a given genus $g$.
Each equivalence class of forms corresponds to a lattice $\Lambda$.
The denominator is the order of the automorphism group $\text{Aut}(\Lambda)$ of the lattice $\Lambda$.
The right-hand-side is the \emph{mass} of the genus of $K$, which is given by a complicated
but explicit formula (see Ref. \onlinecite{Conway88}).

Given a K-matrix for a bosonic state, one can compute the size of its
automorphism group\footnote{For generic K-matrices without any symmetries, the automorphism
group often only consists of two elements: $W=\pm I_{N\times N}$.},
which gives one term in the sum in (\ref{eqn:massformula1}).
If this equals the mass formula on the right-hand-side of Eq. (\ref{eqn:massformula1}),
then it means the genus has only one equivalence class.
If not, we know there is more than one equivalence class in the genus.
Such a program shows \cite{Watson62} that, in fact, all genera contain more than one equivalence class for
$N>10$, i.e. all chiral Abelian quantum Hall states with central charge $c>10$ have multiple
distinct stable chiral edge phases. For $3\leq N \leq 10$, there is a finite set of genera with
only a single equivalence class \cite{Lorch13}; all others have multiple equivalence classes.
The examples of $\nu=16$ analyzed in
Ref.~\onlinecite{Plamadeala13} and $\nu=12/23$ that we gave in Section
\ref{sec:examples} are, in fact, the rule. Bosonic chiral Abelian quantum Hall states with a single
stable chiral edge phase are the exception, they can only exist for $c\leq10$ and they have been completely enumerated\cite{Lorch13}.

This does not tell us how, given one equivalence class, to find other equivalence classes
of $K$-matrices in the same genus. However, one can use the Gauss reduced form \cite{Conway88}
to find all quadratic forms of given rank and determinant by brute force. Then 
we can use the results at the end of previous Section to determine
if the resulting forms are in the same genus.

\subsection{Primary Decomposition of Abelian Topological Phases}

According to the preceding discussion, two distinct edge phases can terminate the
same bulk phase if they are both in the same genus (but not necessarily only if they are in the same genus
in the fermionic case). It may be intuitively clear what this
means, but it is useful to be more precise about what we mean by ``the same bulk phase".
In more physical terms, we would like to be more precise about what it means
for two theories to have the same particle types and $S$- and $T$-matrices.
In more formal terms, we would like to be more precise about what is meant in Nikulin's Theorem 1.11.3
by isomorphic quadratic forms and bilinear forms. In order to do this, it helps
to view an Abelian topological phase in a somewhat more abstract light.
When viewed from the perspective of an edge phase or, equivalently,
a K-matrix, the bulk phase is determined by the signature $({r_+},{r_-})$, together with the bilinear form on the discriminant group $\Lambda^*/\Lambda$ induced by the bilinear form on the dual lattice $\Lambda^*$ determined by $K$. As we have seen, this data uniquely specifies a nondegenerate quadratic form $q \colon \Lambda^*/\Lambda \to \bQ/\bZ$ on the discriminant group.  Therefore, we may view the genus more abstractly in terms of an arbitrary finite Abelian group $A$
and a quadratic form $q \colon A \to \bQ/\bZ$, making no direct reference to an underlying lattice. We will sometimes
call such a quadratic form a \emph{finite} quadratic form to emphasize that its domain
is a finite Abelian group.
The elements of the group $A$ are the particle types in the bulk Abelian topological phase.

Now suppose we have two bulk theories associated with Abelian groups $A$, $A'$,
quadratic forms $q \colon A \to \bQ/\bZ$,  $q'\colon A' \to \bQ/\bZ$
and chiral central charges $c_-$, $c'_-$.
These theories are the same precisely
when the chiral central charges satisfy $c_- \equiv c_-'$  mod 24, and when the associated quadratic forms are \emph{isomorphic}.
This latter condition means that there exists a group isomorphism $f\colon A' \to A$ such that $q' = q \circ f$.
Note that if the quadratic forms are isomorphic then the chiral central charges
must be equal (mod 8) according to the Gauss-Milgram sum.  However,
the bulk theories are the same only if they satisfy the stricter condition that
their central charges are equal modulo 24.

The implications of this become more apparent after observing that any
Abelian group factors as a direct sum $A \simeq \oplus_p A_p$ over primes dividing $|A|$, where $A_p\subset A$ is the \emph{$p$-primary} subgroup of elements with order a power of $p$.   Any isomorphism $f\colon A'\to A$ must respect this factorization by decomposing as $f = \oplus_p f_p$, with each $f_p \colon A_p' \to A_p$.  Furthermore, every finite quadratic form decomposes into a direct sum $q = \oplus_p q_p$ of \emph{$p$-primary} forms; we call $q_p$ the \emph{$p$-part} of $q$.
This ultimately leads to a physical interpretation for $p$-adic integral equivalence: if $p$ is odd, two K-matrices are $p$-adically integrally equivalent precisely when the $p$-parts of their associated quadratic forms are isomorphic.   Additional subtleties arise when $p=2$ but, as we will see, these are the reason for  the distinction between $\sig_x$- and $\sig_z$-equivalence.

The image of a given finite quadratic form $q$ is a finite cyclic subgroup
$N_q^{-1}\bZ/\bZ \subset \bQ/\bZ$
isomorphic to $\bZ/N_q$, where $N_q$ is the \emph{level} of the finite quadratic form $q$.  The level is the smallest integer $N$ such that $q$ factors through $\bZ/N$, implying that the topological spins of particles in $A_q$ are $N_q$th roots of unity.
Because the level of the direct sum of finite quadratic forms is the least common multiple of the levels of the summands, the level of $q = \oplus_p q_p$ is equal to the product $N_q= \prod_p N_{q_p}$ of the levels of the $q_p$.  If $p$ is odd, the level of $q_p$ is the order of the largest cyclic subgroup of $A_p$, while it is typically twice as big for $q_2$.
Physically, this means that the entire theory uniquely factors into a tensor product of anyon theories such that the topological spins of the anyons in the $p$th theory are $p$th-power roots of unity.  This decomposition lets us express a \emph{local-to-global principle} for finite quadratic forms: $q$ and $q'$ are isomorphic iff $q_p$ and $q_p'$ are for every $p$.  Indeed, if one views prime numbers as ``points'' in an abstract topological space\footnote{This space is known as $\Spec(\bZ)$.  Rational numbers are identified with functions on this space according to their prime factorizations.}, this principle says that $q$ and $q'$ are \emph{globally equivalent} (at all primes) iff they are locally equivalent at each prime dividing $|A|$.

Further information about the prime theories is obtained by decomposing each $A_p$ into a product
\begin{equation}
A_p \simeq \prod_{m = 0}^{m_p} (\bZ/p^m)^{d_{p^m}} \label{eqn:Pprimary}
\end{equation}
of cyclic groups, where $d_{p^0}, \dotsc,d_{p^{m_p-1}} \geq 0$ and $d_{p^{m_p}} > 0$.
When $p$ is odd, there is a 1-1 correspondence between bilinear and quadratic forms on $A_p$ because multiplication by 2 is invertible in every $\bZ/p^m$.  Furthermore, given a quadratic form $q_p$ on $A_p$ for odd $p$, we claim there always exists an automorphism $g \in \Aut(A_p)$ that fully diagonalizes $q_p$ relative to a fixed decomposition (\ref{eqn:Pprimary}) such that
\begin{equation}
q_{p}\circ g = \bigoplus_m \big(\underbrace{q_{p^m}^+ \oplus \dotsc \oplus q_{p^m}^+ \oplus q_{p^m}^\pm}_{d_{p^m} \text{ terms}}\big), \label{eqn:diagonal-qp}
\end{equation}
where
\[q^+_{p^m}(x) = \frac{1}{p^m}2^{-1}x^2 \text{ mod } \bZ,\]
\[q^-_{p^m}(x) =  \frac{1}{p^m}u_p2^{-1}x^2 \text{ mod } \bZ\]
and $u_p$ is some fixed non-square modulo $p^n$.
A dual perspective is that,
given $q_p$, it is always possible to choose a decomposition (\ref{eqn:Pprimary}) of $A_p$  relative to which $q_p$ has the form of the right-hand-side of (\ref{eqn:diagonal-qp}).  However, not every decomposition will work for a given $q_p$ because $\Aut(A_p)$ can mix the different cyclic factors.  For example, $\Aut((\bZ/p)^d) \simeq \GL(d,\bZ/p)$ mixes the cyclic factors of order $p$.  There will also be automorphisms mixing lower-order generators with ones of higher order, such as the automorphism of  $\bZ/3\oplus \bZ/9= \vev{\al_3,\al_9}$ defined on generators by $\al_3 \mapsto \al_3$ and $\al_9 \mapsto \al_3 + \al_9$.
Physically, this means that the anyon theory associated to $A_p$ further decomposes into a tensor product of ``cyclic'' theories, although now such decompositions are not unique because one can always redefine the particle types via automorphisms of $A_p$.

\subsection{$p$-adic Symbols}

Two K-matrices are $p$-adically integrally equivalent iff the diagonalizations of the $p$-parts of their associated finite quadratic forms coincide.  The numbers $d_{p^m}$ and the sign of the last form in the $m$th block thus form a complete set of invariants for $p$-adic integral equivalence of K-matrices.  This data is encoded into the \emph{$p$-adic symbol}, which is written as $1^{\pm d_{p^0}} p^{\pm d_{p^1}} (p^2)^{\pm d_{p^2}} \cdots$ (terms with $d_{p^m} = 0$ are omitted) and can be computed using Sage\cite{Sage}.  Two K-matrices are $p$-adically integrally equivalent iff their $p$-adic symbols coincide.

The $p$-adic symbol can be computed more directly by noting that K-matrices are equivalent over the $p$-adic integers when they are equivalent by a \emph{rational} transformation whose determinant and matrix entries do not involve dividing by $p$.  Such transformations can be reduced modulo arbitrary powers of $p$ and give rise to automorphisms of the $p$-part $A_p$ of the discriminant group.  Given a K-matrix $K$, there always exists a  $p$-adically integral transformation $g$ putting $K$ into \emph{$p$-adically block diagonalized}\cite{Conway88} form
\begin{equation}
g K g^T = K_{p^0} \oplus p K_{p^1} \oplus p^2 K_{p^2} \oplus \cdots, \label{eqn:padic_diagonal}
\end{equation}
where $\det(K_{p^m})$ is prime to $p$ for every $m$.

A more direct characterization of the genus can now be given:  Two K-matrices are in the same genus iff they are related by a \emph{rational} transformation whose determinant and matrix entries are relatively prime to twice the determinant, or rather, to the level $N$ of the associated discriminant forms.   Such a transformation suffices to simultaneously $p$-adically block-diagonalize $K$ over the $p$-adic integers for every $p$ dividing twice the determinant, and a similar reduction yields the entire quadratic form on the discriminant group, with some extra complications when $p=2$.  Such a non-integral transformation mapping two edge theories as $g(\Lambda_1) = \Lambda_2$ does not, however induce fractionalization in the bulk since it reduces to an isomorphism between the discriminant groups $\Lambda_1^*/\Lambda_1 \to \Lambda^*_2/\Lambda_2$.
For example, the $\nu = 12/11$ K-matrices (\ref{eqn:twelve-elevenths}) and (\ref{eqn:twelve-elevenths-alt}) are related by the following rational transformation that divides by 3:
\[\pmat{1 & 0 \\ -1/3 & 1}\pmat{3 & 1 \\ 1 & 4} \pmat{1 & -1/3 \\ 0 & 1} = \pmat{1 & 0 \\ 0 & 11}.\]
One might be tempted to look at this transformation and conclude that one of the particle types on the left-hand-side
has undergone fractionalization and divided into $3$ partons (due to the $-1/3$ entries in the matrix), thereby leading to the phase on the right-hand-side. But in mod 11 arithmetic, the number $3$ is invertible, so no fractionalization has
actually occurred.

When $p\neq 2$, the $p$-adic symbol can be directly computed from any such $p$-adic block diagonalization, as the term $(p^m)^{\pm d_{p^m}}$ records the \emph{dimension} $d_{p^m} = \dim(K_{p^m})$ and \emph{sign} $\pm$ of $\det(K_{p^m})$, the latter being given by the \emph{Legendre symbol}
\[\leg{\det(K_{p^m})}{p} = \begin{cases}
+1 & \text{ if } p \text{ is a square mod } p \\
-1 & \text{ if } p \text{ is not a square mod } p.
\end{cases}
\]
In this case, it is further possible to $p$-adically diagonalize all of the blocks $K_{p^m}$, in which case there exists a $p$-adically integral transformation $g$ that diagonalizes the form $Q({\bf x}) = \frac 12 {\bf x}^TK^{-1} {\bf x}$ on the dual lattice $\Lambda^*$ such that its reduction modulo $\Lambda$ takes the form (\ref{eqn:diagonal-qp}).

\begin{table}[htbp]
\centering
\begin{tabular}{|c|lll|c|}
\hline
K-matrix & \multicolumn{3}{|c|}{$p$-adic symbols} & quadratic form \\
\hline
$\smat{1 & 0 \\ 0 & 7}$ & $1^{+2}_0$ & $1^{+1}7^{+1}$ & & \multirow{2}{*}{$q_7^+$} \\
$\smat{2 & 1\\ 1 & 4}$ &  $1^{+2}_\text{even}$ &  $1^{+1}7^{+1}$ &  &\\
\hline
$\smat{1 & 0 \\ 0 & 11}$ & \multirow{2}{*}{$1^{-2}_4$} &  \multirow{2}{*}{$1^{+1}11^{+1}$} & & \multirow{2}{*}{$q_{11}^+$}\\
$\smat{3 & 1\\ 1 & 4}$ & &  & & \\
\hline
$\smat{3 & 0\\ 0 & 5}$ &  $1^{+2}_{0}$ & $1^{-1}3^{+1}$ & $1^{-1}5^{+1}$ &  \multirow{2}{*}{$q_3^+ \oplus q_5^+$}    \\
$\smat{2 & 1\\ 1 & 8}$ &  $1^{+2}_{\text{even}}$ & $1^{-1}3^{+1}$ & $1^{-1}5^{+1}$ & \\
\hline
$\smat{2 & 3\\ 3 & 16}$ &  \multirow{2}{*}{$1^{+2}_{\text{even}}$} & \multirow{2}{*}{$1^{+1}23^{+1}$} & & \multirow{2}{*}{$q_{23}^+$}\\
$\smat{4 & 1\\ 1 & 6}$ &   &  & & \\
\hline
$K_{A_4}$ & $1^{-4}_\text{even}$ & $1^{+3}5^{+1}$ & &  \multirow{2}{*}{$q_{5}^+$} \\
$5\oplus \bI_3$ & $1^{-4}_0$ & $1^{+3}5^{+1}$ & & \\
\hline
$K_{E_8}$ & $1^{+8}_\text{even}$ & & &  \multirow{2}{*}{$0$}  \\
$\mathbb{I}_8$ & $1^{+8}_0$ & & & \\
\hline
$K_{E_8}\oplus \mathbb{I}_4$ & \multirow{3}{*}{$1^{+12}_4$} & & & \multirow{3}{*}{$0$} \\
$\mathbb{I}_{12}$ &  & & &  \\
$K_{D_{12}^+}$  &  & & &  \\
\hline \hline
$\smat{& 2 \\ 2 }$ & $2^{+2}_\text{even}$ & & & $q_{2,2}^+$ \\
\hline
$K_{D_4}$ & $1^{-2}_\text{even}2^{-2}_\text{even}$ & & & $q_{2,2}^-$ \\
\hline
$\smat{4 & 2 \\ 2 & 4}$ & $2^{-2}_\text{even}$ & $1^{+1}3^{+1}$ & & $q_{2,2}^- \oplus q_3^+$ \\
\hline
\end{tabular}
\caption{Here we list the $p$-adic symbols and discriminant quadratic forms for various K-matrices appearing in this paper, beginning with the canonical 2-adic symbol in every case, followed by the symbols for each prime dividing the determinant.  Each block contains inequivalent-but-stably-equivalent matrices.  The last few rows contain K-matrices giving rise to some of the exceptional 2-adic quadratic forms mentioned in the text.
\label{table:padic_symbols}
}
\end{table}

When $p=2$, it is possible that only some of the blocks $K_{2^m}$  in the decomposition (\ref{eqn:padic_diagonal}) can be 2-adically diagonalized\cite{Conway88} (we call these blocks \emph{odd}).  The remaining \emph{even} blocks can only be block diagonalized into $2\times 2$ blocks of the form $\smat{2a & b \\ b & 2c}$ with $b$ odd, or rather, some number of copies of $\sig_x$ and $\smat{2 & 1 \\ 1 & 2}$.   As with odd $p$, the 2-adic symbol associated to such a block diagonalization records the dimensions $d_{2^m}$ of the blocks, together with the \emph{signs} of the determinants $\det(K_{2^m})$, which are given by the \emph{Jacobi symbols}
\[\leg{2}{\det(K_{2^m})} = \begin{cases}
+1 & \text{ if } \det(K_{2^m}) \equiv \pm 1 \text{ mod } 8 \\
-1 & \text{ if } \det(K_{2^m}) \equiv \pm 3 \text{ mod } 8
\end{cases}
\]
and record whether or not $\det(K_{p^m})$ is a square mod 8.  In addition to this data, the 2-adic symbol also records the parities as well as the traces $\Tr K_{2^m}$ mod 8 of the odd blocks.
An additional complication is that a given K-matrix can be 2-adically diagonalized in more than one way, and while the dimensions and parities of the blocks will be the same, the signs and traces of the odd blocks -- and thus the 2-adic symbols -- can be different.  While this makes checking 2-adic equivalence more difficult, it is nonetheless possible to define a \emph{canonical} 2-adic symbol\cite{Conway88} that \emph{is} a complete invariant for 2-adic equivalence.  We record these canonical 2-adic symbols for many of the K-matrices considered in this paper in Table~\ref{table:padic_symbols}.

The reason for the additional complexity when $p=2$ is because multiplication by 2 is not invertible on the 2-primary part $(\bQ/\bZ)_2$ of $\bQ/\bZ$.  This implies that if $q$ refines a bilinear form on a 2-group then so does $q + \frac 12 \text{ mod } \bZ$, and sometimes these refinements are not isomorphic.
For example, there is only one nondegenerate bilinear form $b_2(x,y) = \frac{xy}{2} \text{ mod }  \bZ$ on $\bZ/2$, with two non-isomorphic quadratic refinements $q_2^\pm(x) = \pm \frac x4 \text{ mod }  \bZ$.  Each of these refinements has level 4 and corresponds respectively to the semion $K = (2)$ and its conjugate $K = (-2)$.  These give the $S$ and $T$ matrices
\[S_2=\frac{1}{\sqrt{2}}\pmat{1 & 1 \\ 1 & -1},\,\,\, T_2^\pm = e^{\mp \frac{2\pi i}{24}} \pmat{1 \\ & \pm i}.\]

On $\bZ/2 \times \bZ/2$, there are two isomorphism classes of nondegenerate bilinear forms.  The first class is represented by
\[(b_2 \oplus b_2)(x,y) = \smfrac 12 (x_1 y_1 + x_2 y_2) \text{ mod } \bZ\]
 and has the $S$-matrix
\[S_2 \otimes S_2 =
\frac 12 \pmat{1 & 1 & 1 & 1 \\ 1 & -1 & 1 & -1 \\ 1 & 1 & -1 & -1 \\ 1 & -1 & -1 & 1}.\]
All the refinements in this case have level 4 and are given by tensor products of semions.  Up to isomorphism, this gives three refinements $q_2^+\oplus q_2^+$, $q_2^+\oplus q_2^-$ and $q_2^-\oplus q_2^-$, determined by the K-matrices $\smat{2 \\ & 2}$,  $\smat{2 \\ & -2}$ and  $\smat{-2 \\ & -2}$ with $c_- = 2,0,-2$ respectively.

The second class of bilinear forms on $\bZ/2\times \bZ/2$ contains the single form
\[b_{2,2}(x,y) = \smfrac 12 (x_1 y_2 + x_2 y_1) \text{ mod } \bZ\]
and gives the $S$-matrix
\[S_{2,2} = \frac 12 \pmat{1 & 1 & 1 & 1 \\ 1 & 1 & -1 & -1 \\ 1 & -1 & 1 & -1 \\ 1 & -1 & -1 & 1}.\]
It is refined by two isomorphism classes $q_{2,2}^\pm$ of quadratic forms with T-matrices $T_{2,2}^\pm = \text{diag}(1,\pm 1, \pm 1, -1)$ (these have level 2, the exception to the rule), up to the usual phase of $-2\pi i{c_-}/24$.  The form $q_{2,2}^+$ is given by the K-matrix $\smat{& 2 \\ 2}$ and corresponds to the toric code.  The form $q_{2,2}^-$ is given by the K-matrix
\[K_{D_4} = \pmat{2	&0&	1	&0\\
0	&2	&-1	&0\\
1	&-1	&2	&-1\\
0	&0	&-1	&2}
\]
of $\SO(8)_1$, or equivalently, by the restriction of the quadratic form associated to the K-matrix $\smat{4 & 2 \\ 2 & 4}$ to the 2-part of its discriminant group $\bZ/2\times \bZ/2\times \bZ/3$.  Again, these are distinguished by their signatures, which are 0 and 4 mod 8.  The 2-adic diagonalizations of these K-matrices contain examples of even blocks, as illustrated in to even blocks in Table~\ref{table:padic_symbols}.

Further complexity arises for higher powers of 2:  There are two bilinear forms  $b_4^\pm$ on $\bZ/4$, and four $b_{2^m}^{1,3,5,7}$ on each $\bZ/2^m$ when $m \geq 3$.  There are also four quadratic forms $q_{2^m}^{1,3,5,7}$ on $\bZ/2^m$ for every $m \geq 2$, all with level $2^{m+1}$.  Therefore, the bilinear forms $b_4^{\pm}$ have two refinements each, while the rest have unique refinements.  On top of all this, even more complexity arises from the fact that factorizations of such forms is not typically unique.  It is therefore less straightforward to check equivalence of 2-adic forms.  It is nonetheless still possible to define a canonical 2-adic symbol\cite{Conway88} that is a complete invariant for 2-adic equivalence of K-matrices.  However, this symbol carries strictly more information than the isomorphism class of the 2-part of the discriminant form because it knows the parity of $K$.  To characterize the even-odd equivalences that we investigate in the next section, the usual 2-adic equivalence is replaced with equivalence of the 2-parts of discriminant forms as in the odd $p$ case above.

The 2-adic symbol contains slightly more information than just the equivalence class of a quadratic form on the discriminant group.  This is evident in our even-odd examples, for which all $p$-adic symbols for odd $p$ coincide, with the only difference occurring in the 2-adic symbol.
 It is however clear that two K-matrices $K_\text{even}$ and $K_\text{odd}$ of different parities are stably equivalent precisely when either $K_\text{even} \oplus 1$ and $K_\text{odd} \oplus 1$ are in the same genus, or otherwise, when $K_\text{even} \oplus \sig_z$ and $K_\text{odd} \oplus \sig_z$ are in the same genus.  A detailed study of the 2-adic symbols in this context will appear elsewhere.

\section{Stable Equivalence between Odd and Even Matrices: Fermionic Bulk States
with Bosonic Edges Phases}
\label{sec:odd-even}


We now focus on the case of fermionic systems, which are described by odd $K$-matrices
(i.e., matrices that have at least one odd number on the diagonal).  We ask: Under what
circumstances is such a K-matrix equivalent, upon enlargement by $\sigma_z$
(or $\sigma_x$, since it makes no difference for an odd matrix),
to an even K-matrix enlarged by $\sigma_z$:
\begin{equation}
K_{\rm odd}\oplus\sigma_{z} = W^T (K_{\rm even}\oplus\sigma_z)  W?
\label{eqn:even_odd}
\end{equation}
This question can be answered using the theory of
quadratic refinements.\cite{Belov05,Stirling08}

As we have alluded to earlier, the naive definition of a quadratic form on the discriminant group breaks down for odd matrices.  To be more concrete,
$\frac{1}{2}\vec{u}^2\ ({\rm mod}\ 1)$ is no longer well-defined on the discriminant group. In order to be well-defined on the discriminant group,
shifting $\vec{u}$ by a lattice vector $\bm{\lambda}\in\Lambda$ must leave $q(\vec{u})$ invariant
modulo integers, so that $e^{2\pi iq(\vec{u})}$ in Eq.~(\ref{eqn:T-from-q}) is independent
of which representative in $\Lambda^*$ we take for an equivalence class in $A=\Lambda^*/\Lambda$.
When $K$ is odd, there are some vectors $\bm{\lambda}$ in the original lattice $\Lambda$ such that
\begin{equation}
  {q(\vec{u}+\bm{\lambda})}\equiv {q(\vec{u})+\frac{1}{2}}\,\text{mod } 1.
  \label{}
\end{equation}
Physically, such a vector is just an electron ($\bm{\lambda}\cdot\bm{\lambda}$ is an odd integer). One can attach an odd number of electrons to any quasiparticle and change the exchange statistics by $-1$.
In a sense, the discriminant group should be enlarged to $A\oplus (A+\bm{\lambda}_\text{odd})$: quasiparticles come in doublets composed of particles with opposite fermion parity, and therefore opposite topological twists.
The Gauss-Milgram sum over this enlarged set of quasiparticles is identically zero, which is a clear signature that the Abelian topological phase defined by an odd K-matrix is not a TQFT in the usual sense.

While the $T$ matrix is not well-defined for a fermionic theory, the $S$ matrix, which is determined by the discriminant bilinear form $b([\vec{v}], [\vec{v}'])$, makes perfect sense.
This is because a full braid of one electron around any other particle does not generate a
non-trivial phase.

Given a bilinear form $b$, a systematic approach for defining a quadratic form that is well-defined on the discriminant group comes from the theory of quadratic refinements.
The crucial result is that a given bilinear form can {\it always} be lifted to a quadratic form $q$ on the discriminant group. The precise meaning of ``lifting'' is that there exists a well-defined discriminant quadratic form such that $b([\vec{v}], [\vec{v}'])=q([\vec{v}+\vec{v}'])-q([\vec{v}])-q([\vec{v}'])$.\cite{Belov05,Stirling08}
With $q$, the topological twists are well-defined: $e^{2\pi iq(\vec{u})} = e^{2\pi iq(\vec{u} + \vec{\lambda})}$ for all $\vec{u} \in \Lambda^\ast$ and ${\bf \lambda} \in \Lambda$.
We will give a constructive proof for the existence of such a $q$, given any odd K-matrix.

Once the existence of such a quadratic form $q([\vec{v}])$ is established, we can evaluate the Gauss-Milgram sum (\ref{eqn:Gauss-Milgram}) and determine $c_-$ mod $8$.
We then appeal to the following result of Nikulin~\cite{Nikulin80}:
\medskip

\noindent
Corollary 1.10.2: {\it Given an Abelian group $A$, a quadratic form $q$ on $A$, and positive integers $({r_+},{r_-})$ that satisfy the
Gauss-Milgram sum for $q$, there exists an even lattice with discriminant group $A$, quadratic form $q$ on the discriminant group,
and signature $({r_+},{r_-})$, provided ${r_+}+{r_-}$ is sufficiently-large.}
\medskip

Using Corollary 1.10.2, we immediately see that an even lattice characterized by $(A, q, c_- \text{ mod }8)$ exists,
whose Gram matrix is denoted by $K_\text{even}$.
Recall that the chiral central charge $c_-$ is equal to the signature $\sigma={r_+}-{r_-}$ of the lattice.
Next we show that $K_\text{even}$ is $\sigma_z$-stably equivalent to the odd matrix we started with: namely, \eqref{eqn:even_odd} holds for this $K_\text{even}$.  Since $K_\text{even}$ and $K$ share the same discriminant group and $S$ matrix, they are stably equivalent upon adding unimodular lattices, according to Theorem 1. 1. 9. In other words, there exist unimodular matrices $U$ and $U'$ such that
\begin{equation}
  K\oplus U\simeq K_\text{even}\oplus U'.
  \label{}
\end{equation}
Apparently $U'$ must be odd. We now add to both sides of the equation the conjugate of $U'$ denoted by $\ol{U'}$:
\begin{equation}
  K\oplus (U\oplus \ol{U'})\simeq K_\text{even} \oplus (U'\oplus \ol{U'}).
  \label{}
\end{equation}
On the right-hand side, $U'\otimes \ol{U'}$ is equivalent to $\sigma_z\oplus \sigma_z\oplus \cdots \sigma_z$. On the left-hand side, $U\oplus \ol{U'}$ can be transformed to the direct sum of $\mathbb{I}_{n}$ where $n=\sigma(U)-\sigma(U')=\sigma(K_\text{even})-\sigma(K)$ and several $\sigma_{z/x}$'s. Here $\mathbb{I}_n$ is the $|n|\times |n|$ identity matrix and when $n$ is negative we take it to be $-\mathbb{I}_{|n|}$. If $n\neq 0$ mod $8$, then $K_\text{even}$ has a different chiral central charge as $K$. Therefore we have arrived at the following theorem:
\\ \\
{\it For any odd $K$ matrix,
$K\oplus  \mathbb{I}_{n}$ is $\sigma_z$-stably equivalent to an even K-matrix for an appropriate $n$.
}
\\ \\
The physical implication is that by adding a certain number of Landau levels the edge phase of a fermionic Abelian topological phase is always stably equivalent to a purely bosonic edge phase which has no electron excitations in its low-energy spectrum.

The possible central charges of the bosonic edge theory are $c_{\rm ferm}+n + 8m$
for $m\in \mathbb{Z}$. We can consider a fermionic system with
an additional $8m+n$ Landau levels, where $m$ is the smallest positive integer such that
$8m + n > 0$. Such a fermionic theory has precisely the same discriminant group
as the original fermionic theory and, consequently, is associated with precisely the same bosonic system
defined by the refinement $q([\vec{u}])$. So even if the original fermionic theory does not have a stable
chiral edge phase with only bosonic excitations, there is a closely-related fermionic theory
with some extra filled Landau levels which does have a chiral edge phase whose gapless
excitations are all bosonic. A simple example of this is given by the $\nu=1/5$ Laughlin state,
which has $K=5$. The corresponding bosonic state has $c=4$, so the $\nu=1/5$ Laughlin state
does not have a chiral edge phase whose gapless excitations are all bosonic.
However, the central charges do match if, instead, we consider
the $\nu=3+\frac{1}{5} = 16/5$ state. This state {\it does} have a
bosonic edge phase, with K-matrix
\begin{equation}
K_{A_4}=
\begin{pmatrix}
	2 & 1 & 0 & 0\\
	1 & 2 & 1 & 0\\
	0 & 1 & 2 & 1\\
	0 & 0 & 1 & 2
  \end{pmatrix}
\end{equation}
corresponding to $\SU(5)_1$.  (Ordinarily, the Cartan matrix for $\SU(5)$ is written with $-1$s off-diagonal, but by a change of basis we can make them equal to $+1$.)

In the following we demonstrate concretely how to obtain a particular discriminant quadratic form $q$, starting from the odd lattice given by $K$.  We already know that the naive definition $\frac{1}{2}\vec{u}^2 (\text{mod } 1)$ does not qualify as a discriminant quadratic form.
In order to define a quadratic form on the discriminant group, we first define
a quadratic function ${Q_{\bf w}}(\vec{u})$ according to:
\begin{equation}
  {Q_{\bf w}}(\vec{u})=\frac{1}{2}\vec{u}^2-\frac{1}{2}\vec{u}\cdot\vec{w},
  \label{}
\end{equation}
for $\vec{w} \in \Lambda^\ast$.
Such a linear shift preserves the relation between the
quadratic function ($T$ matrix) and the bilinear form ($S$ matrix):
\begin{equation}
  {Q_{\bf w}}(\vec{u}+\vec{v})-{Q_{\bf w}}(\vec{u})-{Q_{\bf w}}(\vec{v})=\vec{u}\cdot\vec{v}.
  \label{}
\end{equation}
(Notice that $\vec{u}\cdot\vec{v}$ is the symmetric bilinear form $b(\vec{u},\vec{v})$
in Stirling's thesis \cite{Stirling08}). Notice that at this stage ${Q_{\bf w}}$ is not yet a quadratic form on $A$, being just a quadratic function.

If, for any $\bm{\lambda}\in\Lambda$, $Q_{\bf w}$ satisfies ${Q_{\bf w}}(\vec{u}+\bm{\lambda})\equiv {Q_{\bf w}}(\vec{u})\,\text{mod } 1$ or, in other words,
\begin{equation}
  \bm{\lambda}\cdot\bm{\lambda}\equiv \bm{\lambda}\cdot\vec{w}\,\text{mod } 2.
  \label{}
\end{equation}
then we can define the following quadratic form on the discriminant group:
$$
q([\vec{u}]) = {Q_{\bf w}}(\vec{u}).
$$
Expanding $\vec{w}$ in the basis of the dual lattice $\vec{w}=w_I \vec{f}^I$
and expanding $\bm{\lambda}^I \vec{e}_I$, we find that this condition is satisfied
if we take $w_I\equiv K_{II} \,\text{mod } 2$.
Thus, for a Hall state expressed in the symmetric basis, we may identify $\vec{w}$ with twice the spin vector $s_I = K_{II}/2$.\cite{wenzeespin, wenmodular}

A central result of Ref.~\onlinecite{Belov05} is that such a $\vec{w}$
leads to a generalized Gauss-Milgram sum:
\begin{equation}
  \frac{1}{\sqrt{|A|}}e^{\frac{2\pi i}{8}\vec{w}^2}\sum_\vec{u} e^{2\pi i Q_{\vec{w}}(\vec{u})} =
e^{2\pi i\sigma/8},
  \label{}
\end{equation}
where, in order for the notation to coincide, we have replaced the chiral central charge with the signature $\sigma$ on the right-hand-side of the above equation.
Note that the choice of $\vec{w}$ here is not unique.
We can check that the modified Gauss-Milgram sum holds for $\vec{w}+2\bm{\lambda}^*$ where $\bm{\lambda}^*\in\Lambda^*$.
First note that
\begin{multline}
	Q_{\vec{w}+2\bm{\lambda}^*}(\vec{u})=\frac{1}{2}\vec{u}^2-\frac{1}{2}\vec{u}\cdot\vec{w}-\vec{u}\cdot\bm{\lambda}^*\\
	=Q_{\vec{w}}(\vec{u}-\bm{\lambda}^*)-\frac{1}{2}{\bm{\lambda}^*}^2-\frac{1}{2}\bm{\lambda}^*\cdot\vec{w},
  \label{}
\end{multline}
while at the same time
\begin{equation}
  (\vec{w}+2\bm{\lambda}^*)^2 =
  \vec{w}^2+4\bm{\lambda}^*\cdot\vec{w}+4{\bm{\lambda}^*}^2.
  \label{}
\end{equation}
Therefore,
\begin{multline}
  e^{\frac{2\pi i}{8}(\vec{w}+2\bm{\lambda}^*)^2}\sum_\vec{u}e^{2\pi i Q_{\vec{w}+2\bm{\lambda}^*}(\vec{u})}\\ =e^{\frac{2\pi i}{8}\vec{w}^2}\sum_\vec{u}e^{2\pi i Q_{\vec{w}}(\vec{u}-\bm{\lambda}^*)} = e^{2\pi i \sigma/8}.
  \label{}
\end{multline}
One can freely shift $\vec{w}$ by $2\lambda^*$. Consequently, $\vec{w}$ is really
an equivalence class in $\Lambda^*/2\Lambda^*$.


In Appendix~\ref{sec:find-w-in-Lambda}, we further prove that such a representative $\vec{w}$ can always be chosen to lie in the original lattice $\Lambda$.
We denote such a $\vec{w}$ by $\vec{w}_0$.  The advantage of such a choice can be seen from the expression
\[e^{2\pi i Q_{\vec{w}_0}(\vec{u})}=e^{\pi i\vec{u}^2}e^{\pi i \vec{u}\cdot\vec{w}_0}\]
the topological twists. Since $\vec{w}_0$ now lives in $\Lambda$, we have $\vec{u}\cdot\vec{w}_0\in \mathbb{Z}$ and $e^{\pi i\vec{u}\cdot\vec{w}_0}=\pm 1$. This corroborates our intuition that one can salvage the Gauss-Milgram sum in the case of odd matrices by inserting appropriate signs in the sum.

In addition, we can prove that our quadratic function now defines a finite quadratic form because $Q_{\vec{w}_0}(n\vec{u})\equiv n^2 Q_{\vec{w}_0}(\vec{u}) \text{ mod } \bZ$. To see why this is true, we use the definition of $q$:
\begin{eqnarray}
  Q_{\vec{w}_0}(n\vec{u}) &=&\frac{n^2}{2}\vec{u}^2-\frac{n}{2}\vec{u}\cdot\vec{w}_0\cr
&\equiv& \left(\frac{n^2}{2}\vec{u}^2-\frac{n^2}{2}\vec{u}\cdot\vec{w}_0 \right) \text{mod } \mathbb{Z}.
  \label{}
\end{eqnarray}
The second equality follows from the elementary fact that $n^2\equiv n\,(\text{mod } 2)$ together with $\vec{u}\cdot\vec{w}_0\in\mathbb{Z}$. Therefore the definition $q([\vec{u}]) = Q_{\vec{w}_0}(\vec{u})$ mod $\bZ$ is well-defined.


Having found the discriminant quadratic form $q(\vec{u})$, the generalized Gauss-Milgram sum now can be re-interpreted as the ordinary Gauss-Milgram sum of a bosonic Abelian topological phase. As aforementioned, there exists a lifting to an even lattice with the signature $\sigma'\equiv (\sigma-\vec{w}_0^2) \,\text{mod }8$ where $\sigma$ is the signature of the odd matrix $K$ and thus the number of Landau levels we need to add is $n=-\vec{w}_0^2\,\text{ mod }8$.

Hence, we have the sufficient condition for the existence of an even lattice that is stably equivalent
to a given odd lattice: $\sigma'=\sigma$, or $\vec{w}_0^2\equiv 0\,\text{mod } 8$.

An obvious drawback of this discussion is that it is not constructive (which stems from the non-constructive
nature of the proof of Nikulin's theorem \cite{Nikulin80}): we do not know how to construct uniquely the even matrix
corresponding to a given discriminant group, quadratic form $q$,
and central charge $c$.
The distinct ways of lifting usually result in lattices with different signatures.

\section{Novel Chiral Edge Phases of the Conventional Bulk Fermionic
$\nu=8,12, \frac{8}{15}$, $\frac{16}{5}$ states}
\label{sec:examples}

Now that the general framework has been established, in this section we consider a few experimentally relevant examples and their tunneling signatures.

\subsection{$\nu=8$}
\label{sec:n=8}

The integer quantum Hall states are the easiest to produce in experiment and are considered to be well understood theoretically. But surprisingly, integer fillings, too, can exhibit edge phase transitions. The smallest  integer filling for which this can occur is at $\nu = 8$, because eight is the smallest dimension for which there exist two equivalence classes of unimodular matrices. One class contains the identity matrix, $\mathbb{I}_8$, and the other contains $K_{E_8}$, defined in Appendix~\ref{sec:big-matrices}, which is generated by the roots of the Lie algebra of $E_8$. $K_{E_8}$ is an even matrix and hence describes a system whose gapless excitations are all bosonic\cite{Lu12,Plamadeala13} (although if we consider the bosons to be paired fermions, it must contain gapped fermionic excitations.) Yet, counterintuitively, it is stably equivalent to the fermionic $\mathbb{I}_8$; for $W_8$ defined in Appendix~\ref{sec:big-matrices},
\begin{equation} W_8^T(K_{E_8}\oplus \sigma_z)W_8 = \mathbb{I}_8\oplus \sigma_z, \end{equation}
This is an example of the general theory explained in Section \ref{sec:odd-even},
but it is an extreme case in which both phases have only a single particle type -- the trivial particle.
The chiral central charges of both phases are equal and so Nikulin's theorem guarantees that the two bulk phases are equivalent (when the bosonic $E_8$ state is understood to be ultimately built out of electrons) and that there is a corresponding edge phase transition between the two chiral theories.

The action describing the $\mathbb{I}_8$ state with an additional left- and right-moving mode is
\begin{multline}
S = \int dx\,dt \biggl(\frac{1}{4\pi}\left( \mathbb{I}_8 \oplus \sigma_z\right)^{}_{IJ} \partial_t \phi^I \partial_x \phi^J \\ -
\frac{1}{4\pi}V_{IJ}\partial_x \phi^I \partial_x \phi^J 
+ \frac{1}{2\pi}\sum_I \epsilon_{\mu\nu} \partial_\mu \phi^I A_\nu
\biggr).
\end{multline}
The charge vector is implicitly $t_I=1$ for all $I$. As we have shown in previous sections, the basis change $\phi'=W_8\phi$ makes it straightforward to see that if the perturbation
\begin{equation} S' = \int dx dt u' \cos\left( \phi'_9  \pm \phi'_{10}\right)\end{equation}
is the only relevant term, then the two modes $\phi'_9$ and $\phi'_{10}$ would be gapped and the system would effectively be described by $K_{E_8}$.
As in the previous examples, measurements that probe the edge structure can distinguish the two phases of the edge. Consider, first, transport through a QPC that allows tunneling between the two edges of the Hall bar.
In the $\nu=8$ state with $K=\mathbb{I}_8$, the backscattered current will be proportional to the voltage
\begin{equation}
I^b_{\mathbb{I}_8} \propto V
\end{equation}
because the most relevant backscattering operators, $\cos(\phi_I^T - \phi_I^B)$, correspond to the tunneling of electrons.
In contrast, when $K=K_{E_8}$, there is no single-electron backscattering term. Instead, the most relevant operator is the backscattering of charge-$2e$ bosons -- i.e. of pairs of electrons -- from terms like $\cos(\phi_1'^T -\phi_4'^T -\phi_1'^B + \phi_4'^B)$, which yields different current-voltage relation
\begin{equation}
I^b_{E_8} \propto V^3.
\end{equation}

An alternative probe is given by tunneling into the edge from a metallic lead.
In the $K=\mathbb{I}_8$ case, the leading contribution is due to electrons tunneling between the lead and the Hall bar from the terms $\psi^\dagger_{\rm lead}e^{i\phi_I^T}$, yielding
\begin{equation}
I^{\rm tun}_{\mathbb{I}_8} \propto V.
\end{equation}
However, in the $K_{E_8}$ case there are no fermionic charge-$e$ operators to couple to the electrons tunneling from the lead. Instead, the leading term must involve two electrons from the lead tunneling together into the Hall bar. The amplitude for this event may be so small that there is no detectable current. If the amplitude is detectable, then we consider two cases: if the quantum Hall state is not spin-polarized
or if spin is not conserved (e.g. due to spin-orbit interaction), then the leading contribution to the tunneling current is from terms like $\psi_{{\rm lead},\downarrow}^\dagger \psi_{{\rm lead},\uparrow}^\dagger e^{i\phi_1'^T-i\phi_4'^T}$, which represents  two electrons of opposite
spin tunneling together into the Hall bar, yielding
\begin{equation}
I^{\rm tun}_{E_8} \propto V^3.
\end{equation}
If the quantum Hall state is spin-polarized, and tunneling from the lead is spin-conserving,
then the pair of electrons that tunnels from the lead must be a spin-polarized $p$-wave pair, corresponding to a tunneling term like $\psi_{{\rm lead},\downarrow}^\dagger \partial \psi_{{\rm lead},\downarrow}^\dagger e^{i\phi_1'^T-i\phi_4'^T}$ in the Lagrangian,
and we instead expect
\begin{equation}
I^{\rm tun}_{E_8} \propto V^5.
\end{equation}

Another important distinction between the two edge phases is the minimal value of electric charge in the low-energy sector, which can be probed by a shot-noice measurement~\cite{KaneFisherSN, ChamonSN}, as was done in the $\nu=1/3$ fractional quantum Hall state \cite{Picciotto97,Saminadayar97}. The $\mathbb{I}_8$ phase has gapless electrons, so the minimal charge is just the unit charge $e$. However, the $E_8$ edge phase is bosonic and consequently the minimal charge is at least $2e$ (i.e. a pair of electrons). (Electrons are gapped and, therefore, do not contribute to transport at low temperatures and voltages.) Quantum shot noise, generated by weak-backscattering at the QPC is proportional to the minimal current-carrying charge and the average current. So we expect a shot-noise measurement can also distinguish the two edge phases unambiguously.

\subsection{$\nu=12$}
\label{sec:n=12}

In dimensions-9, -10, and -11, there exist two unique positive definite unimodular lattices,
whose $K$-matrices are (in the usual canonical bases)
$\mathbb{I}_{9,10,11}$ or $K_{E_8}\oplus \mathbb{I}_{1,2,3}$.
In each dimension, the two lattices, when enlarged by direct sum with $\sigma_z$,
are related by the similarity transformation of the previous section. However in dimension-12,
a new lattice appears, ${D_{12}^+}$, defined in Appendix~\ref{sec:big-matrices}.
One salient feature of this matrix is that it has an odd element along the diagonal,
but it is not equal to $1$, which is a symptom of the fact that there are
vectors in this lattice that have odd $(\text{length})^2$ but none of them have $(\text{length})^2$=1.
The minimum $(\text{length})^2$ is $2$.
Upon taking the direct sum with $\sigma_z$, the resulting matrix is equivalent
to $\mathbb{I}_{12}\oplus\sigma_z$ -- and hence to $K_{E_8}\oplus\mathbb{I}_4 \oplus \sigma_z$
using the transformation of the previous section -- by the relation $W_{12}^T( K_{D_{12}^+}  \oplus \sigma_z)W_{12} = \mathbb{I}_{12}\oplus \sigma_z$, where $W_{12}$ is defined in Appendix~\ref{sec:big-matrices}.

Consider the action of the $\nu = 12$ state with two additional counter propagating gapless modes and with the implicit charge vector $t_I=1$:
\begin{multline}
S =  \int dx\,dt \biggl(\frac{1}{4\pi}\left(\mathbb{I}_{12} \oplus \sigma_z \right)^{}_{IJ} \partial_t \phi^I \partial_x \phi^J \\ -
\frac{1}{4\pi} V_{IJ}\partial_x \phi^I \partial_x \phi^J
+ \frac{1}{2\pi}\sum_I \epsilon_{\mu\nu} \partial_\mu \phi^I A_\nu
\biggr).
\end{multline}
The matrix $W_{12}$ suggests a natural basis change $\phi'=W_{12}\phi$ in which the perturbation
\begin{equation} S' = \int dx dt u'\cos\left( \phi'_9  \pm \phi'_{10}\right)\end{equation}
can open a gap, leaving behind an effective theory described by $K_{D_{12}^+}$. 


It is difficult to distinguish the $\mathbb{I}_{12}$ edge phase from
the ${E_8}\oplus\mathbb{I}_4$ phase because both phases have charge-$e$ fermions
with scaling dimension-$1/2$. However, both of these edge phases can be distinguished
from the $D_{12}^+$ phase in the manner described for the $\nu=8$ phases
in the previous subsection. At a QPC, the most relevant backscattering terms will have scaling dimension 1; one example is the term $\cos(\phi_{11}'^T-\phi_{11}'^B)$, which yields the current-voltage relation
\begin{equation}
I^b_{D_{12}^+} \propto V^3.
\end{equation}
This is the same as in the $E_8$ edge phase at $\nu=8$ because the most-relevant
backscattering operator is a charge-$2e$ bosonic operator with scaling dimension $2$.
There is a charge-$e$ fermionic operator $\exp(i(\phi_2'^T+2\phi_{12}'^T))$,
but it has scaling dimension $3/2$.
Its contribution to the backscattered current is
$\propto V^5$, which is sub-leading compared to the contribution above, although its
bare coefficient may be larger. However, if we couple the edge to a metallic lead via $\psi_{\rm lead}^\dagger \exp(i(\phi_2'^T+2\phi_{12}'^T))$,
single-electron tunneling is the dominant contribution for a spin-polarized edge, yielding
\begin{equation}
\label{eqn:D12+tunneling}
I^{\rm tun}_{D_{12}^+} \propto V^3,
\end{equation}
while pair tunneling via the coupling $\psi_{\rm lead}^\dagger \partial \psi_{\rm lead}^\dagger e^{i\phi_{11}'^T}$ gives a sub-leading contribution $\propto V^5$. If the edge is spin-unpolarized,  pair tunneling via the coupling $\psi_{\rm lead, \uparrow}^\dagger \psi_{\rm lead,\downarrow}^\dagger e^{i\phi_{11}'^T}$ gives a contribution with the same $V$ dependence as single-electron tunneling.

\subsection{Fractional Quantum Hall States with Multiple Edge Phases}
\label{sec:fractional}

In Section \ref{sec:illustrative}, we discussed the $\nu=8/7$ state, which has two possible edge phases.
Our second fermionic fractional quantum Hall example is
\begin{equation}
\label{eqn:eight-fifteenths}
  K_1=
  \begin{pmatrix}
	3 & 0 \\
	0 & 5
  \end{pmatrix}
\end{equation}
with $t=(1,1)^T$. We again assume that a pair of gapped modes
interacts with these two modes, and we assume that they are modes of oppositely-charged particles (e.g. holes), so that $t=(1,1,-1,-1)^T$.
Upon enlarging by $\sigma_z$, we find that $K_1 \oplus \sigma_z = W^T (K_2 \oplus\sigma_z ) W$,
where
\begin{equation}
\label{eqn:eight-fifteenths-alt}
  K_2=
  \begin{pmatrix}
	2 & 1 \\
	1 & 8
  \end{pmatrix}
\end{equation}
and
\begin{equation}
\label{eqn:eight-fifteenths-basis}
  W =
  \begin{pmatrix}
	1 & 3 & 0 & 1\\
	0 & 3 & 0 & 1\\
    0 & 0 & 1 & 0\\
    1 & 8 & 0 & 3\\
  \end{pmatrix}.
\end{equation}
If the following perturbation is relevant, it gaps out a pair of modes:
\begin{equation}
S' = \int dx \, dt \, u' \cos(-3{\phi_1}-5{\phi_2}+{\phi_3}+3{\phi_4}).
\end{equation}
Under the basis change (\ref{eqn:eight-fifteenths-basis}),
$-3{\phi_1}-5{\phi_2}+{\phi_3}+3{\phi_4} = \phi_3' + \phi_4'$, so the remaining theory
has K-matrix (\ref{eqn:eight-fifteenths-alt}).

In the $K_1$ edge phase (\ref{eqn:eight-fifteenths}), the backscattered current at a QPC is dominated by the tunneling term $\cos(\phi_2^T - \phi_2^B)$, which yields
\begin{equation}
I_1^b \propto V^{-3/5},
\end{equation}
while the tunneling current from a metallic lead is dominated by the single-electron tunneling term $\psi_{\rm lead}^\dagger e^{3i\phi_1^T}$, which yields
\begin{equation}
I_1^{\rm tun} \propto V^3.
\end{equation}
In the $K_2$ edge phase (\ref{eqn:eight-fifteenths-alt}), the backscattered current at a QPC is dominated by the tunneling term $\cos(\phi_2'^T - \phi_2'^B)$, yielding
\begin{equation}
I_1^b \propto V^{-11/15},
\end{equation}
while the tunneling current from a metallic lead is dominated by the pair-tunneling term $\psi_{\rm lead}^\dagger\partial\psi_{\rm lead}^\dagger e^{i\phi_1'^T -7i\phi_2'^T}$, which assumes a spin-polarized edge, and yields
\begin{equation} I^\text{tun}_2 \propto V^{11}. \end{equation}

As we discussed in Section~\ref{sec:odd-even}, the $\nu=16/5$ state can have two possible edge phases,
one with
\begin{equation}
K_1 =
\begin{pmatrix}
	1 & 0 & 0 & 0\\
	0 & 1 & 0 & 0\\
	0 & 0 & 1 & 0\\
	0 & 0 & 0 & 5
  \end{pmatrix},
\end{equation}
which is essentially the edge of the $\nu=1/5$ state, together with $3$ integer quantum Hall edges.
The other possible phase has
\begin{equation}
K_2 =
\begin{pmatrix}
	2 & 1 & 0 & 0\\
	1 & 2 & 1 & 0\\
	0 & 1 & 2 & 1\\
	0 & 0 & 1 & 2
  \end{pmatrix}.\label{eq:k-16-5-even}
\end{equation}
Upon enlarging by a pair of gapped modes, the two matrices are related by $K_1\oplus \sigma_z = W^T(K_2 \oplus \sigma_z)W$, where
\begin{equation}
W=
\left(
\begin{array}{cccccc}
 1 & 0 & 0 & 2 & 0 & -1 \\
 -1 & 1 & 0 & -4 & 0 & 2 \\
 1 & -1 & 1 & 6 & 0 & -3 \\
 -1 & 1 & -1 & -8 & 1 & 4 \\
 0 & 0 & 0 & 5 & 0 & -2 \\
 -1 & 1 & -1 & -10 & 1 & 5
\end{array}
\right)
\end{equation}
If the gapped modes are oppositely charged holes, then the following perturbation carries no charge:
\begin{equation} S' = \int dx dt u'\cos(-\phi_1+\phi_2-\phi_3-5\phi_4+\phi_5+3\phi_6) \end{equation}
If this perturbation is relevant, it will gap out a pair of modes and leave behind an effective theory describe by the K-matrix (\ref{eq:k-16-5-even}),

The two edge phases of the $\nu=16/5$ state can be distinguished by the voltage dependence of the current backscattered at a quantum point contact and the tunneling current from a metallic lead. In the $K_1$ edge phase, the backscattered current at a QPC is dominated by the quasiparticle backscattering term $\cos(\phi_4^T - \phi_4^B)$, yielding the current-voltage relation
\begin{equation}
  I^b_1\propto V^{-3/5}. 
  \label{}
\end{equation}
In the $K_2$ edge phase, there are several terms that are equally most-relevant, including, for example $\cos(\phi_1'^T-\phi_1'^B)$, which yield the current-voltage relation
\begin{equation}
 I^b_2\propto V^{3/5}. 
  \label{}
\end{equation}
Meanwhile, in the $K_1$ edge phase, single-electron tunneling from a metallic lead given by, for example, $\psi_{\rm lead}^\dagger e^{i\phi_1^T}$, yields the dependence
\begin{equation}
  I^{\rm tun}_1\propto V, 
  \label{}
\end{equation}
while in the $K_2$ edge phase there are only pair-tunneling terms; one such term for a spin-polarized edge is $\psi_{\rm lead}^\dagger \partial \psi_{\rm lead}^\dagger e^{i\phi_1'^T + i\phi_4'^T}$, which yields
\begin{equation}
  I^{\rm tun}_2\propto V^{5}.
  \label{}
\end{equation}

We now consider an example of a bosonic fractional quantum Hall state with $\nu=12/23$,
\begin{equation}
\label{eqn:twelve-twenty-thirds}
  K^b_1 =
  \begin{pmatrix}
	2 & 3 \\
	3 & 16
  \end{pmatrix}
\end{equation}
and $t=(1, 1)^T$. (This is a natural choice of charge vector for bosonic atoms in a rotating trap. For paired electrons in a magnetic field, it would be more natural to have $t=(2, 2)^T$) By a construction similar to the one discussed
in the fermionic cases of $\nu=8, 12, 8/7, 8/15$ and the bosonic integer quantum Hall cases of $\nu=8, 16$,
this state has another edge phase described by
\begin{equation}
\label{eqn:twelve-twenty-thirds-alt}
  K^b_2 =
  \begin{pmatrix}
	4 & 1 \\
	1 & 6
  \end{pmatrix}
\end{equation}
and $t=(1,-1)^T$. As in the previous cases, the two edge phases
can be distinguished by transport through a QPC or tunneling from a metallic lead.

\section{Some Remarks on Genera of Lattices and Bulk Topological Phases}
\label{sec:remarks}

The focus in this paper is on the multiple possible gapless edge phases associated with a given bulk topological phase.
However, having established that the former correspond to lattices while the latter correspond to genera of lattices (or, possibly, pairs of genera of lattices),
we note here that some results on genera of lattices published by Nikulin in Ref. \onlinecite{Nikulin80}
have direct implications for bulk topological phases.
We hope to explore these relations more thoroughly in the future.

We begin by noting that the data that determine a genus of lattices is precisely the data that determine
a $2+1$-D Abelian topological phase.
Recall that the elements of the discriminant group $A$ of a lattice form the particle content of an Abelian topological phase.
We can turn this around by noting that the particle content and fusion rules of any Abelian topological phase can
be summarized by an Abelian group $A$ whose elements are the particle types in the theory and whose
multiplication rules give the fusion rules of the theory. The fusion rules take the form of
the multiplication rules of an Abelian group because only one term can appear on the right-hand-side of the fusion rules
in an Abelian topological phase. Meanwhile, specifying the $S$-matrix for the topological phase is equivalent to giving
a bilinear form on the Abelian group $A$ according to $S_{{[\bf v]},{[\bf v']}} =\frac{1}{\sqrt{|A|}} e^{-2\pi i b([{\bf v}],[{\bf v'}])}$.
A quadratic form $q$ on the Abelian group $A$ determines the topological twist factors or, equivalently, the $T$-matrix
of an Abelian topological phase according to $\theta_{[{\bf v}]}  = e^{2\pi i q([\vec{v}])}$. Finally, the signature of
the form, the number of positive and negative eigenvalues $r_+$ and $r_-$ of the quadratic form $q$, determines the right
and left central charges, according to $c_R = r_+$ and $c_L=r_-$. The chiral central charge $c_- = c_R - c_L$ is given by
${c_-}={r_+}-{r_-}$ which, in turn, determines the modular transformation properties of states and, consequently, the partition functions of the bulk
theory on closed $3$-manifolds (e.g. obtained by cutting a torus out of $S^3$, performing a Dehn twist, and gluing it back in).
The signature is determined (mod 8) by the quadratic form $q$, according to
the Gauss-Milgram sum:
$$
 \frac{1}{\sqrt{|A|}}\sum_{a\in A}e^{2\pi iq(a)}=e^{2\pi i c_-/8}
$$
We now consider Nikulin's Theorem 1.11.3, given in Section \ref{sec:stable} and also his result
\medskip

\noindent
Proposition 1.11.4: {\it There are at most $4$ possible values for the signature} (mod 8) {\it for the quadratic forms associated with a given bilinear form
on the discriminant group.}
\\

Theorem 1.11.3 (given in Section \ref{sec:stable}) states that the $S$-matrix and ${r_+}-{r_-}$ (mod 8) completely and uniquely
determine the $T$-matrix, up to relabellings of the particles that leave the theory invariant.
In Section \ref{sec:odd-even} we show constructively that such a $T$-matrix exists in the fermionic case.
Proposition 1.11.4 tells us that, for a given $S$-matrix, there are at most
$4$ possible values for the signature ${r_+}-{r_-}$ (mod 8) and, therefore, at most $4$ possible
$T$-matrices. One way to interpret this is that the elements of the $T$-matrix are the square roots of the diagonal elements of
the $S$-matrix; therefore, they can be determined, up to signs from the $S$-matrix. There are, at most, four consistent ways of
doing this, corresponding to, at most, four possible values of the Gauss-Milgram sum.

Then, Theorem 1.10.2, stated in Section \ref{sec:odd-even},
tells us that the quadratic form defines an even lattice. Thus, to any fermionic Abelian topological phase,
we can associate a bosonic Abelian
topological phase with the same particle types, fusion rules, and $S$-matrix.
The bosonic phase has a well-defined $T$-matrix, unlike the fermionic phase.
In addition, we have:
\medskip

\noindent
Theorem 1.3.1: {\it Two lattices $S_1$ and $S_2$ have isomorphic bilinear forms on their discriminant groups if and only if
there exist unimodular lattices $L_1$, $L_2$ such that $S_1 \oplus L_1 \cong S_2 \oplus L_2$.}
\\

In other words, two lattices have isomorphic bilinear forms if they
are stably equivalent under direct sum with arbitrary unimodular lattices, i.e. if we are allowed to take direct sums with
arbitrary direct sums of $\sigma_x$, $\sigma_z$, $1$, and $K_{E_8}$.
One example of this is two lattices in the same genus. They have the same parity, signature, and bilinear form and
are stably equivalent under direct sum with $\sigma_x$, as required by the theorem. However, we can also consider lattices
that are not in the same genus. The example that is relevant to the present discussion is a pair of theories, one of which is fermionic and the other
bosonic. They have the same $S$-matrix but may not have the same chiral central charges. The theorem tells us that the difference can be made
up with unimodular theories. But since $\sigma_x$ and $\sigma_z$ do not change the chiral central charge, the unimodular lattices
given by the theorem must be hypercubic lattices. (In the fermionic context, the $E_8$ lattice is $\sigma_z$-stably equivalent to the $8$-dimensional
hypercubic lattice.) In other words, \emph{every fermionic Abelian topological phase is equivalent to a bosonic Abelian topological phase, together
with some number of filled Landau levels}.

Finally, we consider Nikulin's Corollary 1.16.3, given in Section \ref{sec:stable}, which states that
the genus of a lattice is determined by its parity, signature, and bilinear form on the discriminant group.
Recall that the parity of a lattice is even or odd according whether its K-matrix is even or odd.
The even case can occur in a purely bosonic system while the odd case necessarily requires ``fundamental" fermions, i.e. fermions
that braid trivially with respect to all other particles.
Therefore, specifying the parity, signature, and bilinear form on an Abelian group $A$ is equivalent to specifying (1) whether or not
the phase can occur in a system in which the microscopic constituents are all bosons, (2) the
$S$-matrix, and (3) the chiral central charge. (According to the previous theorem, the $T$-matrix is determined by
the latter two.) This is sufficient to specify any Abelian topological phase.
According to Corollary 1.16.3, these quantities specify a genus of lattices. Thus, given any Abelian topological phase,
there is an associated genus of lattices. We can take any lattice in this genus,
compute the associated K-matrix (in some basis) and define a $\U(1)^{{r_+}+{r_-}}$
Chern-Simons theory. A change of basis of the lattice corresponds to a change of variables in the Chern-Simons theory. Different lattices in the same genus
correspond to different equivalent $\U(1)^{{r_+}+{r_-}}$ Chern-Simons theories for the same topological phase.
Therefore, it follows from Corollary 1.16.3 that {\it every Abelian topological phase can be represented as a $\U(1)^{N}$ Chern-Simons theory}.

\section{Discussion}
\label{sec:discussion}


A theoretical construction of a bulk quantum Hall
state typically suggests a particular edge phase, which we will call $K_1$.
The simplest example of this is given by integer quantum
Hall states, as we discussed in Sections \ref{sec:preliminaries} and \ref{sec:examples}.
However, there is no reason to believe that the state observed in experiments is in this
particular edge phase $K_1$.
This is particularly important because the exponents associated with gapless edge excitations, as measured through quantum point contacts, for instance, are among the few ways to identify the topological order of the state \cite{Wen90a,Nayak08}. In fact, such experiments are virtually the only way to probe the state
in the absence of interferometry experiments \cite{Chamon97,Fradkin98,Bonderson06a,Stern06,Willett09a,Willett10a,Willett13a}
that could measure quasiparticle braiding properties. Thus, given an edge theory $K_2$
that is deduced from experiments, we need to know if a purely edge phase transition
can take the system from $K_1$ to $K_2$ -- in other words, whether the edge theory $K_2$
is consistent with the proposed theoretical construction of the bulk state.
We would also like to predict, given an edge theory $K_2$ deduced from experiments,
what other edge phases ${K_3}, {K_4}, \ldots$ might be reached by tuning parameters at the edge,
such as the steepness of the confining potential.
In this paper, we have given answers to these two questions.

The exotic edge phases at $\nu=8, 12$ discussed in this paper may be realized
in experiments in a number of materials which display the integer quantum Hall effect.
These include Si-MOSFETs \cite{vonKlitzing80}, GaAs heterojunctions and quantum wells (see, e.g. Refs. \onlinecite{Prange90},
\onlinecite{DasSarma97} and references therein),
InAs quanutm wells \cite{Brosiga98}, graphene \cite{Young12},
polar ZnO/Mg$_{x}$Zn$_{1-x}$O interfaces \cite{Tsukazaki07}. In all of these systems, edge excitations can interact strongly
and could be in an $E_8$ phase at $\nu=8$ or the $D_{12}^+$ phase or the
${E_8}\oplus \mathbb{I}_4$ phase at $\nu=12$. To the best of our knowledge, there
are no published studies of the detailed properties of edge excitations at these integer quantum
Hall states.

The novel edge phase that we have predicted at $\nu=16/5$ could occur at the
$\nu=3+1/5$ state that has been observed\cite{Eisenstein02}
in a 31 million cm$^2$/Vs mobility GaAs quantum well. This edge phase is dramatically
different than the edge of the $\nu=1/5$ Laughlin state weakly-coupled to
$3$ filled Landau levels.
Meanwhile, a $\nu=8/15$ state could occur in an unbalanced double-layer system (or, possibly, in a single wide quantum well) with $\nu=1/3$ and $1/5$ fractional quantum Hall states
in the two layers. Even if the bulks of the two layers are
very weakly-correlated, the edges may interact strongly, thereby leading to the alternative edge
phase that we predict. Finally, if an $\nu=8/7$ state is observed,
then, as in the two cases mentioned above, it could have an edge phase without gapless fermionic excitations.

We have focussed on the relationship between the $K$-matrices of different edge phases of the same bulk.
However, in a quantum Hall state, there is also a $t$-vector, which specifies how the topological phase is coupled
to the electromagnetic field. An Abelian topological phase specified by a $K$-matrix splits into several phases
with inequivalent $t$-vectors. Therefore, two different $K$-matrices that are stably equivalent may still belong to different
phases if the corresponding $t$-vectors are are not related by the appropriate similarity transformation.
However, in all of the examples that we have studied, given a $(K,t)$ pair, and a $K'$ stably equivalent to $K$, we were always
able to find a $t'$ related to $t$ by the appropriate similarity transformation. Said differently, we were always able to find an edge phase transition
driven by a charge-conserving perturbation. It would be interesting to see if there are cases in which there is no charge-conserving
phase transition between stably-equivalent $K$, $K'$ so that charge-conservation symmetry presents an obstruction
to an edge phase transition between $K$, $K'$.

When a bulk topological phase has two different edge phases, one that supports gapless fermionic excitations
and one that doesn't, as is the case in the $\nu=8$ integer quantum Hall state and the
fractional states mentioned in the previous paragraph, then a domain wall at the edge must support a fermionic zero mode.
For the sake of concreteness, let us consider the $\nu=8$ IQH edge.
Suppose that the edge of the system lies along the $x$-axis and the edge is in the conventional phase
with $K=\mathbb{I}_8$ for $x<0$ and the $K_{E_8}$ phase for $x>0$.
The gapless excitations of the edge are fully chiral; let us take their chirality to be such that
they are all right-moving. A low-energy fermionic excitation propagating along
the edge cannot pass the origin since there are no gapless fermionic excitations in the $E_8$ phase.
But since the edge is chiral, it cannot be reflected either. Therefore, there must be a fermionic zero mode
at the origin that absorbs it.

We discussed how the quadratic refinement allows us to relate a given fermionic theory to a bosonic one. One example that we considered in detail related $K_1 = \begin{pmatrix} 1 & 0 \cr 0 & 7 \end{pmatrix}$ to $K_2 = \begin{pmatrix} 2 & 1 \cr 1 & 4 \end{pmatrix}$. Both of these states are purely chiral. However, we noted that we are not restricted to relating purely chiral theories; we could have instead considered a transition between the $\nu = 1/7$ Laughlin edge and the non-chiral theory described by $K = \begin{pmatrix} 2 & 1 & 0 \cr 1 & 4 & 0 \cr 0 & 0 & -1 \end{pmatrix}$. This transition does not preserve chirality, but the chiral central charges of the two edge theories are the same. It can be shown that there exist regions in parameter space where the non-chiral theory is stable -- for example, if the interaction matrix, that we often write as $V$, is diagonal, then the lowest dimension backscattering operator has dimension equal to $4$. Even more tantalizingly, it is also possible to consider the $\nu = 1/3$ Laughlin edge which admits an edge transition to the theory described by $K' = \begin{pmatrix} - 2 & - 1 \cr -1 & -2 \end{pmatrix} \oplus \mathbb{I}_{3 \times 3}$. The upper left block is simply the conjugate or $(-1)$ times the Cartan matrix for $\SU(3)_1$. About the diagonal $V$ matrix point, the lowest dimension backscattering term is marginal; it would be interesting to know if stable regions exist.

The theory of quadratic refinements implies that any fermionic TQFT can be realized as a bosonic one, together with some filled Landau levels,
as we discussed as the end of Sec. \ref{sec:remarks}. In particular, it suggests the following picture: a system of fermions forms a
weakly-paired state in which the phase of the complex pairing function winds $2N$ times around the Fermi surface.
The pairs then condense in a bosonic topological phase.
The winding of the pairing function gives the additional central charge (and, if the fermions are charged, the same Hall conductance) as $N$ filled
Landau levels. The remarkable result that follows from the theory of quadratic refinements is that all Abelian fermionic topological phases
can be realized in this way.

In this paper, we have focused exclusively on fully chiral states. However, there are many
quantum Hall states that are not fully chiral, such as the $\nu=2/3$ states. The stable edge
phases of such states correspond to lattices of indefinite signature.
Once again, bulk phases of bosonic systems correspond to genera of lattices
while bulk phases of fermionic systems correspond either to genera of lattices
or to pairs of genera -- one even and one odd. Single-lattice genera are much more common
in the indefinite case than in the definite case \cite{Conway88}. If an $n$-dimensional genus has more than one lattice in
it then $4^{[\frac{n}{2}]}d$ is divisible by $k^{n \choose 2}$ for some non-square natural number $k$
satisfying $k \equiv 0 \mbox{ or } 1 \pmod 4$, where $d$ is the determinant of the associated Gram matrix (i.e. the K-matrix). In particular, genera containing multiple equivalence classes of $K$-matrices
must have determinant greater than or equal to $17$ if their rank is $2$; greater than
or equal to $128$ if their rank is $3$; and $5^{n \choose 2}$ or $2\cdot 5^{n \choose 2}$
for, respectively, even or odd rank $n\geq 4$.

Quantum Hall states are just one realization of topological phases. Our results apply to other
realizations of Abelian topological states as well. In those physical realizations which do not
have a conserved $\U(1)$ charge (which is electric charge in the quantum Hall case),
there will be additional $\U(1)$-violating operators which could tune the edge of a system between different phases.

Although we have, in this paper, focussed on Abelian quantum Hall states, we believe that
non-Abelian states can also have multiple chiral edge phases. This will occur
when two different edge conformal field theories with the same chiral central charge are associated with
the same modular tensor category of the bulk. The physical mechanism underlying
the transitions between different edge phases associated
with the same bulk is likely to be the same as the one discussed here.
In this general case, we will not be able to
use results on lattices and quadratic forms to find such one-to-many bulk-edge correspondances.
Finding analogous criteria would be useful for interpreting experiments on the $\nu=5/2$ fractional
quantum Hall state.

\acknowledgements
We would like to thank Maissam Barkeshli, Andrei Bernevig, Parsa Bonderson, Michael Freedman,
Taylor Hughes, Chenjie Wang, and Zhenghan Wang for helpful discussions. We thank David Clarke
for discussions and for sharing unpublished work \cite{Clarke-unpublished} with us.
C.N. and E.P. have been partially supported by the DARPA QuEST
program. C.N. has been partially supported by the AFOSR under grant FA9550-10-1-0524.
J.Y. has been partially supported by the NSF under grant 116143.  J.C. acknowledges the support of the National Science Foundation Graduate Research Fellowship under Grant No. DGE1144085.

\appendix

\section{A Non-Trivial Example of using the Gauss-Smith Normal Form to find the
Discriminant Group}
\label{sec:finding-discriminant}

We now apply the method described in Section \ref{sec:stable}
to the $\SO(8)_1$ theory,which is  given by the following $K$ matrix:
\begin{equation}
K=
\begin{pmatrix}
2	&0&	1	&0\\
0	&2	&-1	&0\\
1	&-1	&2	&-1\\
0	&0	&-1	&2
\end{pmatrix}
\end{equation}
It is not clear, simply by inspection, what vectors correspond to generators of the fusion group.

The Gauss-Smith normal form is
\begin{equation}
D=
\begin{pmatrix}
1	&0	&0&	0\\
0	&1	&0	&0\\
0	&0	&2	&0\\
0	&0	&0	&2
\end{pmatrix}
\end{equation}
Hence, the fusion group of the theory is $\mathbb{Z}/2 \times \mathbb{Z}/2$.

and the Q matrix
\begin{equation}
Q=
\begin{pmatrix}
2	&0	&1	&0\\
3	&1	&0	&1\\
2	&0&	0	&1\\
1	&0	&0	&0
\end{pmatrix}
\end{equation}
So the fusion group is generated by the two quasiparticles corresponding to $(2, 0, 0, 1)$ and $(1, 0, 0, 0)$. We can then compute the $S$, $T$
matrices and the result agrees with what is known (all nontrivial quasiparticles are fermions and they have semionic mutual braiding statistics with each other).

Another useful piece of information from the Smith normal form is that the discriminant group for a $2\times 2$
K-matrix
\begin{equation}
K =
\begin{pmatrix}
a & b\\
b & c
\end{pmatrix}
\end{equation}
with $\gcd(a, b, c)=1$  and $d = |ac-b^2|$ is $\mathbb{Z}/d$. More generally, it is $\mathbb{Z}/f \times \mathbb{Z}/(d/f)$ when $\gcd(a,b,c) = f$.

\section{Proof that $\vec{w} \in \Lambda$ exists such that $\bm{\lambda}\cdot\bm{\lambda}\equiv\bm{\lambda}\cdot \vec{w} \text{ mod }2$ for all $\bm{\lambda} \in \Lambda$}
\label{sec:find-w-in-Lambda}


We begin by showing that for any K-matrix, there exists a set of integers $w_J$ such that
 \begin{equation} K_{II} \equiv \sum_{J=1}^N K_{IJ}w_J \text{ mod }2, \text{ for all }I \label{eq:wlinear}\end{equation}
 where $N$ is the dimension of the K-matrix.

 Assume the K-matrix has $M\leq N$ rows that are linearly independent mod 2; denote these rows $R_1, ... R_M$ and define the set $R=\lbrace R_i \rbrace$. The linear independence of the $R_i$ implies that Eq~(\ref{eq:wlinear}) is satisfied for these rows, i.e., there exists a set of integers $(w_0)_J$ satisfying
 \begin{equation} K_{II} \equiv \sum_{J=1}^N K_{IJ}(w_0)_J \text{ mod }2, \text{ for all }I\in R\label{eq:wlinearR}\end{equation}
 For a row $I \not\in R$, the elements of the $I^\text{th}$ row in $K$ can be written as a linear combination of the rows in $R$:
 \begin{equation} K_{IJ} \equiv \sum_{R_i\in R} c_{IR_i}K_{R_iJ} \text{ mod }2, \text{ for }I\not\in B \end{equation}
 where the $c_{IR_i} \in \lbrace 0,1 \rbrace$ are coefficients. It follows that for $I\not\in R$:
 \begin{align} K_{II}&\equiv \sum_{R_i\in R}c_{IR_i}K_{R_iI} \equiv  \sum_{R_i\in R}c_{IR_i}K_{IR_i} \nonumber\\
 &\equiv \sum_{R_i,R_j\in R}c_{IR_i}c_{IR_j}K_{R_iR_j} \equiv \sum_{R_i\in R}c_{IR_i}^2K_{R_iR_i}\nonumber\\
 & \equiv \sum_{R_i\in R}c_{IR_i}K_{R_iR_i} \text{ mod }2
 \end{align}
Furthermore, for $I\not \in R$
\begin{align}  \sum_{J=1}^N K_{I J}(w_0)_J & \equiv \sum_{J=1}^N \sum_{R_i\in R}c_{IR_i}K_{R_iJ} (w_0)_J \nonumber \\
&\equiv \sum_{R_i\in R}c_{IR_i}K_{R_iR_i} \text{ mod }2\end{align}
Hence, for $I\not\in R$, $K_{II} \equiv \sum_{J=1}^N K_{IJ}(w_0)_J \text{ mod }2 $. Since this equation already holds for $I\in R$, we have shown that $w_0$ is a solution to Eq~(\ref{eq:wlinear}).

It follows that for any choice of $\bm{\lambda}=\lambda_J \vec{e}_J \in \Lambda$,
\begin{align} \bm{\lambda} \cdot \bm{\lambda} &= \sum_{I,J=1}^N \lambda_I\lambda_JK_{IJ} \equiv \sum_{I=1}^N\lambda_I K_{II}\nonumber\\
  &\equiv \sum_{I=1}^N \lambda_I \sum_{J=1}^N K_{IJ}(w_0)_J \equiv \bm{\lambda}\cdot \bm{w}_0 \text{ mod }2
\end{align}
where $\vec{w}_0 = (w_0)_J \vec{e}_J$ is a vector in $\Lambda$.

\onecolumngrid

\section{Relevant large matrices}
\label{sec:big-matrices}

Here we define matrices referred to in \ref{sec:examples}:

\begin{equation}
K_{E_8} = \begin{pmatrix}
	2 & -1 & 0 & 0 & 0 & 0 & 0 & 0  \\
	-1 & 2 & -1 & 0 & 0 & 0 & -1 & 0 \\
	0 & -1 & 2 & -1 & 0 & 0 & 0 & 0 \\
	0 & 0 & -1 & 2 & -1 & 0 & 0 & 0 \\
	0 & 0 & 0 & -1 & 2 & -1 & 0 & 0 \\
	0 & 0 & 0 & 0 & -1 & 2 & 0 & 0 \\
	0 & -1 & 0 & 0 & 0 & 0 & 2 & -1 \\
	0 & 0 & 0 & 0 & 0 & 0 & -1 & 2 \\	
	\end{pmatrix}
\end{equation}

\begin{equation} W_8 = \begin{pmatrix}
-5 & -5 & -5 & 5 & 5 & 5 & 5 & 5 & 8 & 16\\
-10 & -10 & -10 & 9 & 9 & 9 & 9 & 9 & 15 & 30\\
-8 & -8 & -8 & 8 & 7 & 7 & 7 & 7 & 12 & 24\\
-6 & -6 & -6 & 6 & 6 & 5 & 5 & 5 & 9 & 18\\
-4 & -4 & -4 & 4 & 4 & 4 & 3 & 3 & 6 & 12\\
-2 & -2 & -2 & 2 & 2 & 2 & 2 & 1 & 3 & 6\\
-7 & -7 & -6 & 6 & 6 & 6 & 6 & 6 & 10 & 20\\
-4 & -3 & -3 & 3 & 3 & 3 & 3 & 3 & 5 & 10\\
1 & 1 & 1 & -1 & -1 & -1 & -1 & -1 & -3 & -4\\
-2 & -2 & -2 & 2 & 2 & 2 & 2 & 2 & 4 & 7
\end{pmatrix}
\end{equation}


\begin{equation}
K_{D_{12}^+} = \left( \begin{array}{cccccccccccc}
2 & 0 & 1 & 0 & 0 & 0 & 0 & 0 & 0 & 0 & 0 & -1\\
0 &  2 &  -1 &  0 &  0 &  0 &  0 &  0 &  0 &  0 &  0 &  0 \\
1 &  -1 &  2 &  -1 &  0 &  0 &  0 &  0 &  0 &  0 &  0 &  0 \\
0 &  0 &  -1 &  2 &  -1 &  0 &  0 &  0 &  0 &  0 &  0 &  0 \\
0 &  0 &  0 &  -1 &  2 &  -1 &  0 &  0 &  0 &  0 &  0 &  0 \\
0 &  0 &  0 &  0 &  -1 &  2 &  -1 &  0 &  0 &  0 &  0 &  0 \\
0 &  0 &  0 &  0 &  0 &  -1 &  2 &  -1 &  0 &  0 &  0 &  0 \\
0 &  0 &  0 &  0 &  0 &  0 &  -1 &  2 &  -1 &  0 &  0 &  0 \\
0 &  0 &  0 &  0 &  0 &  0 &  0 &  -1 &  2 &  -1 &  0 &  0 \\
0 &  0 &  0 &  0 &  0 &  0 &  0 &  0 &  -1 &  2 &  -1 &  0 \\
0 &  0 &  0 &  0 &  0 &  0 &  0 &  0 &  0 &  -1 &  2 &  0 \\
-1 &  0 &  0 &  0 &  0 &  0 &  0 &  0 &  0 &  0 &  0 &  3
\end{array} \right)
\end{equation}

\begin{equation}
W_{12}=
\left(
\begin{array}{cccccccccccccc}
 11 & 6 & 6 & -6 & -6 & -6 & -6 & -6 & -6 & -6 & -6 & -6 & 0 & 22 \\
 -9 & -4 & -5 & 5 & 5 & 5 & 5 & 5 & 5 & 5 & 5 & 5 & 0 & 18 \\
 -18 & -9 & -9 & 10 & 10 & 10 & 10 & 10 & 10 & 10 & 10 & 10 & 0 & 36 \\
 -16 & -8 & -8 & 8 & 9 & 9 & 9 & 9 & 9 & 9 & 9 & 9 & 0 & 32 \\
 -14 & -7 & -7 & 7 & 7 & 8 & 8 & 8 & 8 & 8 & 8 & 8 & 0 & 28 \\
 -12 & -6 & -6 & 6 & 6 & 6 & 7 & 7 & 7 & 7 & 7 & 7 & 0 & 24 \\
 -10 & -5 & -5 & 5 & 5 & 5 & 5 & 6 & 6 & 6 & 6 & 6 & 0 & 20 \\
 -8 & -4 & -4 & 4 & 4 & 4 & 4 & 4 & 5 & 5 & 5 & 5 & 0 & 16 \\
 -6 & -3 & -3 & 3 & 3 & 3 & 3 & 3 & 3 & 4 & 4 & 4 & 0 & 12 \\
 -4 & -2 & -2 & 2 & 2 & 2 & 2 & 2 & 2 & 2 & 3 & 3 & 0 & 8 \\
 -2 & -1 & -1 & 1 & 1 & 1 & 1 & 1 & 1 & 1 & 1 & 2 & 0 & 4 \\
 3 & 2 & 2 & -2 & -2 & -2 & -2 & -2 & -2 & -2 & -2 & -2 & 0 & -7 \\
 0 & 0 & 0 & 0 & 0 & 0 & 0 & 0 & 0 & 0 & 0 & 0 & 1 & 0 \\
 2 & 1 & 1 & -1 & -1 & -1 & -1 & -1 & -1 & -1 & -1 & -1 & 0 & -4
 \end{array} \right)
\end{equation}


\twocolumngrid

\bibliography{qh-edge-phases}

\end{document}